\documentclass[12pt,12pt]{article}
\usepackage{amsmath,amssymb}
\usepackage{slashed} 

\catcode`\@=11
\newdimen\ex@
\def\beq{\begin{equation}}
\def\eeq{\end{equation}}
\def\beqa{\begin{eqnarray}}
\def\eeqa{\end{eqnarray}}
\newcommand{\ba}{\begin{eqnarray}}
\newcommand{\ea}{\end{eqnarray}}
\newcommand\BA{\begin{array}}
\newcommand\EA{\end{array}}

\begin{document}
\def\thefootnote{\fnsymbol{footnote}}

\title{\bf
Anomaly-Free
Supersymmetric $\frac{SO(2N\!+\!2)}{U(N\!+\!1)}$\\
$\sigma$-Model Based on the $SO(\!2N\!\!+\!\!1\!)$ Lie Algebra\\
of the Fermion Operators\footnotemark[1]}
\vskip0.4cm
\author{Seiya NISHIYAMA$\!$\footnotemark[2]~,~
Jo\~ao da PROVID\^{E}NCIA$\!$\footnotemark[3]\\ 
Constan\c{c}a PROVID\^{E}NCIA\footnotemark[4]~~and
Fl\' avio CORDEIRO\footnotemark[5]\\
\\[-0.4cm]
Centro de F\'\i sica Computacional,
Departamento de F\'\i sica,\\
Universidade de Coimbra,
P-3004-516 Coimbra, Portugal
\\ \\[-0.4cm]
{\it Dedicated to the Memory of Hideo Fukutome}}

\maketitle

\vskip0.35cm

\footnotetext[1]
{A preliminary version of
this work has been presented by S. Nishiyama
at the YITP Workshop \\
~~~~~YITP-W-09-04 on
{\it Development of Quantum Fields Theory and String Theory 2009},\\
~~~~~Yukawa Institute for Theoretical Physics, Kyoto University, Kyoto, Japan, 6-10 July, 2009.}
\footnotetext[2]
{Corresponding author.
~E-mail address: seikoceu@khe.biglobe.ne.jp,~
nisiyama@teor.fis.uc.pt}
\footnotetext[3]
{E-mail address: providencia@teor.fis.uc.pt}
\footnotetext[4]
{E-mail address: cp@teor.fis.uc.pt}
\footnotetext[5]
{E-mail address: flaviocordeiro\_704@hotmail.com}

\vspace{-1.5cm}

\begin{abstract}
The extended supersymmetric (SUSY) $\sigma\!$-model has been proposed on
the bases of $SO(2N \!\!+\!\! 1)$ Lie algebra spanned by fermion
annihilation-creation operators and pair operators.
The canonical transformation,
extension of an $SO(2N)$ Bogoliubov transformation
to an $SO(2N \!\!+\!\! 1)$ group, is introduced.
Embedding the $SO(2N \!\!+\!\! 1)$ group into
an $SO(2N \!\!+\!\! 2)$ group and
using $SO(2N \!\!+\! \!2)/U(N \!\!+\!\! 1)$ coset variables,
we have investigated the SUSY $\sigma$-model on the K\"{a}hler manifold,
the coset space
$SO(2N \!\!+\!\! 2)/U(N \!\!+\!\! 1)$.
We have constructed the Killing potential, extension of
the potential in the $SO(2N)/U(N)$ coset space
to that in the $SO(2N \!\!+\!\! 2)/U(N \!\!+\!\! 1)$ coset space.
It is equivalent to the generalized density matrix
whose diagonal-block part is related to a reduced scalar potential
with a Fayet-Ilipoulos term.
The $f$-deformed reduced scalar potential is optimized with respect to
vacuum expectation value
of the $\sigma$-model fields and a solution for one of the $SO(2N \!\!+\!\! 1)$
group parameters has been obtained.
The solution,
however,
is only a small part of all solutions
obtained from anomaly-free SUSY coset models.
To construct the coset models consistently,
we must embed a coset coordinate in an anomaly-free spinor representation (rep)
of $SO(2N \!\!+\!\! 2)$ group
and give corresponding K\"{a}hler and Killing potentials for an anomaly-free
$SO(2N \!\!+\!\! 2)/U(N \!\!+\!\! 1)$ model
based on each positive chiral spinor rep.
Using such mathematical manipulation
we construct successfully
the anomaly-free $SO(2N \!\!+\!\! 2)/U(N \!\!+\!\! 1)$
SUSY $\sigma$-model
and investigate new aspects
which have never been seen
in the SUSY $\sigma$-model on the K\"{a}hler coset space
$SO(2N)/U(N)$.
We reach
a $f$-deformed reduced scalar potential.
It is minimized
with respect to the vacuum expectation value
of anomaly-free SUSY $\sigma$-model fields.
Thus 
we find an interesting $f\!$-deformed solution
very different from the previous solution
for an anomaly-free
$SO(2 \!\cdot\! {\bf 5} \!\!+\!\! 2)/(SU({\bf 5} \!\!+\!\! 1) \!\times\! U(1))$
SUSY $\sigma$-model.
\end{abstract}
\vspace{-0.2cm}
~~~~~~~~$\!$PACS
11.10.Lm, 12.60.Jv


\newpage

\setcounter{equation}{0}
\renewcommand{\theequation}{\arabic{section}.\arabic{equation}}

\section{Introduction}

~~
The supersymmetric (SUSY) extension of the nonlinear $\sigma$-model
was first given by Zumino under the introduction of scalar fields
\cite{Zumino.79}
which take values in a complex K\"{a}hler manifold.
The extended $\sigma$-model defined on symmetric spaces
have been intensively studied
in modern elementary particle physics,
superstring theory and supergravity theory
\cite{NNH.01}.
The $\sigma$-model on the hyper K\"{a}hler manifold
also has been deeply investigated in various contexts
\cite{LRHKLR.8387}.

The Hartree-Bogoliubov theory (HBT)
\cite{Bog.59}
has been regarded as the standard approximation
in the theory of fermion systems
\cite{RS.80}.
In the HBT
an HB wave function for such systems represents
a Bose condensate of fermion pairs.
Standing on the Lie-algebraic viewpoint,
the fermion pair operators form an $SO(2N)$ Lie algebra
and contain a $U(N)$ Lie algebra as a subalgebra
where $N$ denotes the number of fermion states.
The $SO(2N)( \!=\! g)$ and $U(N)( \!=\! h)$ mean a special orthogonal group
of $2N$ dimensions and a unitary group of $N$ dimensions, respectively.
One can give an integral representation of a state vector
under the group $g$,
the exact coherent state representation (CS rep) of a fermion system
\cite{Perelomov.86}.

A procedure for consistent coupling of gauge- and matter superfields to SUSY $\sigma$-models on the K\"{a}hler coset spaces
has been given by van Holten et al.
These authors have presented a mathematical tool of constructing
a Killing potential and have applied their method
to the explcit construction of SUSY $\sigma$-models
on the coset spaces $\frac{SO(2N)}{U(N)}$.
They have shown that
only a finite number of the coset models can be consistent
when coupled to matter superfields with $U(N)$ quantum numbers
reflecting spinor reps of $SO(2N)$
\cite{NNH.01}.
Deldug and Valent have investigated the K\"{a}hlerian $\sigma$-models
in two-dimensional space-time at the classical and quantum levels.
They have presented a unified treatment of the models
based on irreducible hermitian symmetric spaces 
corresponding to the coset spaces $\frac{G}{H}$
\cite{DelVal.85}.
On the other hand,
van Holten et al. and Higasijima et al.
have also discussed the construction of $\sigma$-models
on compact and non-compact Grassmannian manifolds,
$\frac{SU(N+M)}{S[U(N) \times U(M)]}$ and
$\frac{SU(N,M)}{S[U(N) \times U(M)]}$
\cite{vanHolten.85,HKN.02}.

Fukutome et al. have proposed
a new fermion many-body theory
based on the $SO(2N \!\!+\!\! 1)$ Lie algebra of fermion operators
composed of single annihilation $c_{\alpha }$ and creation $c^{\dag }_{\alpha }$ operator $(\alpha \!=\! 1,\dots,N)$
and pair operator
\cite{FYN.77}.
A rep of an $SO(2N \!\!+\!\! 1)$ group has been derived
by a group extension of the $SO(2N)$ Bogoliubov transformation
for fermions
to a new canonical transformation group.
The fermion Lie operators, when operating on
the integral rep of
the $SO(2N \!\!+\!\! 1)$ wave function, are mapped into
the regular rep of the $SO(2N \!\!+\!\! 1)$ group and
are represented by boson operators.
Bosonization of creation-annihilation and pair operators
is given in
\cite{YN.76}.

Along the same strategy
we have proposed an extended SUSY $\sigma$-model
on K\"{a}hler coset space
$\!\frac{G}{H} \!=\!\! \frac{SO(2N \!+\! 2)}{U(N \!+\! 1)}\!$
based on the $SO(2N \!\!+\!\! 1)$ Lie algebra of the fermion operators
\cite{SJCF.08}
(referred to as I).
Embedding the $SO(2N \!\!+\!\! 1)$ group into an $SO(2N \!\!+\!\! 2)$ group and
using
$\!\frac{SO(2N \!+\! 2)}{U(N \!+\! 1)}\!$ coset variables
\cite{Fuk.77},
we have studied a new aspect of the SUSY $\sigma$-model on
the K\"{a}hler manifold of the coset space
$\!\frac{SO(2N \!+\! 2)}{U(N \!+\! 1)}\!$.
If we introduce annihilation and creation operators,
$c_{\!N\!+\!1}$ and $c^\dag _{\!N\!+\!1}$,
for a fictitious degree of freedom $N\!+\!1$,
an unphysical space of $2^{N+1}$ dimensions is inevitably met.
Then the $\frac{SO(2N\!+\!2)}{U(N\!+\!1)}$ coset description
under the algebra $SO(2N\!\!+\!\!2)$ is essential and important.
In this context we have constructed a Killing potential.
It is greatly surprising that the Killing potential
is equivalent to the generalized density matrix in the HBT.
Its diagonal-block part is related to a reduced scalar potential
with a Fayet-Ilipoulos term.
After rescaling the Goldstone fields by a parameter $f$
(inverse of mass $m_\sigma$),
minimization of the $f$-deformed reduced scalar potential has led us to an interesting solution of the SUSY $\sigma$-mdoel.

The solution of
the $\frac{SO(2N \!+\! 2)}{U(N \!+\! 1)}$ SUSY $\sigma$-model,
however, 
is only a small part of all solutions
obtained from anomaly-free SUSY coset models.
A consistent theory of coupling of gauge- and matter-superfields to
SUSY $\sigma$-model has been proposed on the K\"{a}hler coset space.
As shown by van Holten et al.
\cite{NNH.01},
if we construct some quantum field theories
based on pure coset models,
we meet with a serious problem of anomalies
in a holonomy group
which particularly occur in pure SUSY coset models
due to the existence of chiral fermions.
Using a coset coordinate in an anomaly-free spinor rep of the $SO(2N)$ group,
these authors have constructed a Killing potential and
have applied their method to the explicit construction of the SUSY
$\sigma$-model on the coset space $\frac{SO(2N)}{U(N)}$.
For analysis of the anomaly see Ref.
\cite{Bert.96}.
The anomaly cancellation condition was first given
by Georgi and Glashow
\cite{GG.72}.
The Adler-Bell-Jackiw anomaly
often occurs in gauged-SUSY nonlinear $\sigma$-models
jointly with $\sigma$ fermions
\cite{Ong.83,MN.84}.
The $\sigma$-models are based on the K\"{a}hler manifolds
$\frac{G}{H}$ ($H$: Subgroup of $G$).
di Vecchia et al. stated that
some nonlinear $\sigma$-models in which the scalar manifold is
a coset space $\frac{G}{H}$ may show up anomalies when
they are coupled to fermions which are not
in anomaly-free reps of the subgroup $H$ of $G$
\cite{VFG.85}.
Anomalies in compact and non-compact K\"{a}hler manifolds
have been intensively discussed
\cite{CG.85,AoyamavanHolten.85}.
This is also the case for our orthogonal coset
$\frac{SO(2N\!+\!2)}{U(N\!+\!1)}$,
though a spinor rep of the $SO(2N\!+\!2)$ group is anomaly free.
To construct a consistent SUSY coset model,
we must embed a coset coordinate
in an anomaly-free spinor rep of $SO(2N\!+\!2)$ group
and give corresponding K\"{a}hler and
Killing potential for the anomaly-free
$\frac{SO(2N\!+\!2)}{U(N\!+\!1)}$ model
based on a positive chiral spinor rep.
To achieve such an object in the case of $SO(2N)$ group/algebra,
van Holten et al. have proposed
a method of constructing
the K\"{a}hler and Killing potentials
\cite{NNH.01}.
This idea is very suggestive and useful
for our present aim of constructing
the corresponding K\"{a}hler and Killing potentials
for the case of $SO(2N\!+\!2)$ group/algebra.

The $\frac{SO(10)}{U(5)}$ coset model is proposed in the Standard Model
and constructed by the $SU(5) \!\times\! U(1)$ fermionic fields content
of one generation of quarks and leptons, including a right-handed neutrino.
There is, however, no coset model such as $\frac{SO(12)}{U(6)}$ in the Standard Model.
Thus we choose the $\frac{SO(10)}{U(5)}$ coset model as a basic model
and extend it to an
$\frac{SO(2\cdot{\bf 5}\!+\!2)}{SU({\bf 5}\!+\!1)\times U(1)}$ SUSY $\sigma$-model.

In Section 2,
we give a brief summary
of an $SO(2N \!\!+\!\! 1)$ canonical transformation,
embedding of $SO(2N \!\!+\!\! 1)$ group into an $SO(2N \!\!+\!\! 2)$ one and fixing a
$\!\frac{SO(2N \!+\! 2)}{U(N \!+\! 1)}$ coset variable.
In Section 3,
we recapitulate a SUSY $\sigma$-model
on the coset space $\frac{SO(2N \!+\! 2)}{U(N \!+\! 1)}$
and its Lagrangian on the K\"{a}hler manifold,
the symmetric space $\frac{SO(2N \!+\! 2)}{U(N \!+\! 1)}$.
The theory is invariant under a SUSY transformation
and the Killing potential is expressed in terms of the coset variables.
If gauge fields
are introduced in the model,
the theory becomes no longer invariant under the transformation.
To restore the SUSY property, it is inevitable to introduce gauginos,
auxiliary fields and Fayet-Ilipoulos terms, which make
the theory invariant under the SUSY transformation,
i.e., chiral invariant and produces a $f$-deformed reduced scalar potential.
Optimization of the $f$-deformed reduced scalar potential reproduces
the $f$-deformed solution in I.
In Section 4,
we show that the optimized $f$-deformed solution
satisfies the idempotency relation
$
< \!\!\! W \!\!\! >_{\!f\mbox{\scriptsize min}}^2
=
< \!\!\! W \!\!\! >_{\!f\mbox{\scriptsize min}}
$
for a factorized density matrix
$< \!\!\! W \!\!\! >_{\!f\mbox{\scriptsize min}}$.
We also present a vacuum function for bosonized fermions
in terms of the Nambu-Goldstone condensate and $U\!(1)$ phase. 
In Section 5,
we construct
an anomaly-free $\frac{SO(2N\!+\!2)}{U(N\!+\!1)}$ SUSY $\sigma$-model
and see what subjects are new.
We give an invariant Killing potential
which is exactly derived for $SU(N\!+\!1)$ tensors.
In Section 6,
we give a new $f$-deformed reduced scalar potential.
After optimization of the $f$-deformed reduced scalar potential,
we find an interesting $f$-deformed solution
for an anomaly-free
$\frac{SO(2\cdot{\bf 5} \!+\! 2)}{SU({\bf 5}\!+\!1) \!\times\! U(1)}$ SUSY $\sigma$-model
which has a very different aspect from the previous solution.
Finally in Section 7,
we give discussions and some concluding remarks.


\newpage

\setcounter{equation}{0}
\renewcommand{\theequation}{\arabic{section}.\arabic{equation}}

\section{Brief summary of embedding of $SO(2N\!+\!1)$ Bogoliubov transformation into $SO(2N\!+\!2)$ group}

\def\bra#1{{<\!#1\,|}} 
\def\ket#1{{|\,#1\!>}}

~~~~
Following I
we give a brief summary
of the $SO(2N \!\!+\!\! 1)$ canonical transformation,
of embedding of $SO(2N \!\!+\!\! 1)$ group into an $SO(2N \!\!+\!\! 2)$ group
and of fixing a $\!\frac{SO(2N \!+\! 2)}{U(N \!+\! 1)}$ coset variable.
Let $c_{\alpha }$ and $c^{\dag }_{\alpha }$, $\alpha$ 
\!=\! 
$1,\dots,N$, be fermion annihilation and creation operators
satisfying the canonical anti-commutation relations
$
\{c_{\alpha },c^{\dag }_{\beta }\}
\!=\!
\delta_{\alpha \beta }
~\mbox{and}~
\{c^{\dag }_{\alpha },c^{\dag }_{\beta }\}
\!=\!
\{c_{\alpha },c_{\beta }\}
\!=\! 0
$.
The set of fermion operators
$
c_{\alpha },c^{\dag }_{\alpha } ,~
E^{\alpha }_{~\beta }
\!=\!
c^{\dag }_{\alpha }c_{\beta }
\!-\! 1/2 \!\cdot\! \delta_{\alpha \beta } ,~
E^{\alpha \beta }
\!=\!
c^{\dag }_{\alpha }c^{\dag }_{\beta }
$
and
$
E_{\alpha \beta }
\!=\!
c_{\alpha }c_{\beta }
$ 
form an $SO(2N \!+\! 1)$ Lie algebra.
The $SO(2N \!+\! 1)$ Lie algebra of the fermion operators contains
the $U(N)(\!=\! \{E^{\alpha }_{~\beta }\})$ and
the $SO(2N)(\!=\! \{E^{\alpha }_{~\beta },
~E^{\alpha \beta },~E_{\alpha \beta }\})$
Lie algebras of the pair operators 
as subalgebras.

An $SO(2N)$ canonical transformation $U(g)$ belongs to 
the fermion $SO(2N)$ Lie operators.
The transformation $U(g)$ is the generalized Bogoliubov transformation 
\cite{Bog.59} 
specified by an $SO(2N)$ matrix $g$\\[-12pt]
\beq
U(g)(c, c^{\dag })U^{\dag }(g)
\!=\!
(c, c^{\dag }) g ,~~~
g^{\dag }g \!=\! gg^{\dag } \!=\! 1_{2N} ,~~
\det g
\!=\!
1 ,
\label{Bogotrans}
\eeq
\beq
U(g)U(g') \!=\! U(gg') ,~~~
U(g^{-1}) \!=\! U^{-1}(g) \!=\! U^{\dag }(g) ,~~~
U(1_{2N}) \!=\! \mathbb{I} .
\label{Ug}
\eeq
($c$, $c^{\dag }$) is
a 2$N$-dimensional row vector
(($c_{\alpha }$), ($c^{\dag }_{\alpha }$)).
$a \!=\! (\!a^{\alpha }_{~\beta })$ and $b \!=\! (\!b_{\alpha \beta })$
are $N \!\times\! N$ matrices.
The HB ($SO(2N)$) wave function $\ket g$ is generated as
$\ket g \!=\! U(g) \ket 0$ 
($\ket 0$ : the vacuum satisfying 
$c_{\alpha }\ket 0 \!=\! 0$).
The wave function $\ket g$ is expressed as\\[-20pt]
\beqa
\ket g
\!=\!
\bra 0 U(g) \ket 0
\exp(1 / 2 \cdot q_{\alpha \beta }c^\dagger_\alpha 
c^\dagger_\beta) \ket 0 ,
\label{Bogoketg}
\eeqa
\vspace{-1.0cm}
\beqa
\bra 0 U(g) \ket 0
\!=\!
\overline{\Phi }_{00}(g)
\!=\!
\left[\det(a)\right]^{1/2}
\!=\!
\left[\det(1_N + q^\dag q)\right]^{-1/4}
e^{i \tau / 2}~,
\label{Bogowf}
\eeqa
\vspace{-1.0cm}
\beqa
q \!=\! ba^{-1} \!=\! -q^{\mbox{\scriptsize T}},
\mbox{(variable of the $SO(2N) / U(N)$ coset space)} ,
\tau
\!=\!
i / 2 \ln \!
\left[\det({a}^*) / \! \det({a})
\right] .
\label{Bogocoset}
\eeqa\\[-20pt]
The symbols $\det$ and {\scriptsize T} denote
the determinant and transposition, respectively.
The overline denotes the complex conjugation.

The canonical anti-commutation relation gives us not only
the above two Lie algebras but also a third algebra.
Let $n$ be the fermion number operator
$n \!=\! c^\dag _{\alpha } c_\alpha$.
The operator $(-1)^n$ anticommutes with
$c_\alpha$ and $c^\dag _\alpha$;
$
\{ c_\alpha,(-1)^n \}
\!=\!
\{ c^\dag _\alpha,(-1)^n \}
\!=\!
0
$.
Let us introduce an operator
$
\Theta
$
as
$
\Theta
\!=\!
\theta_\alpha c^\dag_\alpha \!-\! \overline{\theta }_\alpha c_\alpha
$.
Here we use the summation convention over repeated indices.
Due to the relation
$
\Theta ^2
\!=\!
-
\overline{\theta }_\alpha \theta_\alpha
$,
we have
\beqa
\left.
\BA{ll}
e^\Theta
\!=\!
Z 
\!+\! X_\alpha c^\dag_\alpha
\!-\! \overline{X}_\alpha c_\alpha ,~~
\overline{X}_\alpha X_\alpha \!+\! Z^2 \!=\! 1 ,\\
\\[-8pt]
Z
=
\cos \theta ,~~
X_\alpha
\!=\!
\theta_\alpha / \theta \sin \theta ,~~
\theta ^2
\!=\!
\overline{\theta }_\alpha \theta_\alpha .
\EA
\right\}
\label{theta}
\eeqa
From
the anti-commutator of $(-1)^n$ with
$c_\alpha$ and $c^\dag _\alpha$ and (\ref{theta}),
we obtain
\beqa
~
e^\Theta (c_\alpha, c^\dag _\alpha ,
{\displaystyle \frac{1}{\sqrt{2}}}) (-1)^n e^{-\Theta }
\!=\!
(c_\beta, c^\dag _\beta ,
{\displaystyle \frac{1}{\sqrt{2}}}) (-1)^n \!
\left[ \!\!
\BA{ccc} 
\delta_{\beta \alpha } 
\!-\! 
\bar{X}_\beta X_\alpha &
\bar{X}_\beta \bar{X}_\alpha & -\sqrt{2}Z \bar{X}_\beta \\
\\[-8pt]
X_\beta X_\alpha & \delta_{\beta \alpha } 
\!-\! 
X_\beta \bar{X}_\alpha & 
\sqrt{2}ZX_\beta  \\
\\[-8pt]
\sqrt{2}ZX_\alpha & -\sqrt{2}Z \bar{X}_\alpha & 2Z^2 \!-\! 1
\EA \!\!
\right] .
\label{chiraloptrans}
\eeqa
From
(\ref{Bogotrans}), (\ref{chiraloptrans})
and the commutator of $U(g)$ with $(-1)^n$,
we obtain\\[-16pt]
\beqa
U(G)(c_\alpha, c^{\dag }_\alpha ,1 / \sqrt{2}) (-1)^n U^{\dag }(G)
\!=\!
(c_\beta, c^{\dag }_\beta, 1 / \sqrt{2}) (-1)^n \!
\left[ \!
\BA{ccc} 
A_{\beta \alpha } & \overline{B}_{\beta \alpha } &
-\overline{x}_\beta / \sqrt{2} \\
\\[-10pt]
B_{\beta \alpha } & \overline{A}_{\beta \alpha } & 
x_\beta / \sqrt{2} \\
y_\alpha / \sqrt{2} & 
-\overline{y}_\alpha / \sqrt{2} & z
\EA \!
\right] ,
\label{SO2Nplus1chiraltrans}
\eeqa
\vspace{-0.5cm}
\beqa
\!\!\!\!\!\!
\left.
\BA{ll}
&
A_{\alpha \beta }
\!=\!
a_{\alpha \beta }
\!-\!
\overline{X}_\alpha Y_\beta
\!=\!
a_{\alpha \beta }
\!-\!
\overline{x}_\alpha y_\beta / 2(1 + z) ,~~
B_{\alpha \beta }
\!=\!
b_{\alpha \beta }
\!+\!
X_\alpha Y_\beta
\!=\!
b_{\alpha \beta }
\!+\!
x_\alpha y_\beta  / 2(1 \!+\! z) ,\\
\\[-4pt]
&
x_\alpha
\!=\!
2ZX_\alpha ,~
y_\alpha
\!=\!
2ZY_\alpha ,~
z
\!=\!
2Z^2 \!-\! 1 ,~
Y_\alpha
\!=\!
X_\beta a^\beta_{~\alpha } 
- 
\bar{X}_\beta b_{\beta \alpha }.
\EA
\right\}
\label{relAtoaXY}
\eeqa
Equation
(\ref{SO2Nplus1chiraltrans})
can be written as\\[-10pt]
\beq
U(G)(c, c^{\dag },1 / \sqrt{2}) U^{\dag }(G)
\!=\!
(c, c^{\dag }, 1 / \sqrt{2}) 
(z \!-\! \rho)G ,~
{\rho }
\!=\!
x_{\alpha }c^{\dag }_{\alpha }
-
\overline{x}_{\alpha }c_{\alpha } ,~
{\rho }^2 
\!=\!
- \overline{x}_{\alpha }x_{\alpha } \!=\! z^2 \!-\! 1 ,
\eeq
\beq
G 
\!\stackrel{\mathrm{def}}{=}\!
\left[ \!
\BA{ccc} 
A & \overline{B} & -\overline{x} / \sqrt{2} \\
\\[-10pt]
B & \overline{A} & x / \sqrt{2} \\
y / \sqrt{2} &
-\overline{y} / \sqrt{2} & z
\EA \!
\right],
~
G^{\dag }G \!=\! GG^{\dag } \!=\! 1_{2N \!+\! 1} ,~
\det G
\!=\!
1 ,
\eeq
\beq
U(G)U(G') \!=\! U(GG') ,~
U(G^{-1}) \!=\! U^{-1}(G) \!=\! U^{\dag }(G) ,~
U(1_{2N \!+\! 1}) \!=\! \mathbb{I}_G .
\eeq
($c$, $c^{\dag }$, $1 / \sqrt{2}$) is
a (2$N \!+\! 1$)-dimensional row vector
(($c_{\alpha }$), ($c^{\dag }_{\alpha }$), $1 / \sqrt{2}$).
$A \!=\! (A^{\alpha }_{~\beta })$ and $B \!=\! (B_{\alpha \beta })$
are $N \!\times\! N$ matrices.
The transformation $U(G)$ is a nonlinear transformation
with a gauge factor $z \!-\! \rho$
\cite{FYN.77}.
The $SO(2N \!\!+\!\! 1)$ canonical transformation $U(G)$ is generated by
the fermion $SO(2N \!\!+\!\! 1)$ Lie operators.
The transformation $U(G)$ is an extension of
the generalized Bogoliubov transformation $U(g)$
\cite{Bog.59} 
to a nonlinear transformation and is specified
by the $SO(2N \!\!+\!\! 1)$ matrix $G$.

The $SO(2N \!+\! 1)$ wave function
\cite{Fuk.77,Fuk.81}
$\ket G \!=\! U(G) \ket 0$
is expressed as
\beqa
\ket G
\!=\!
\bra 0 U(G) \ket 0 (1 \!+\! r_\alpha c^\dagger_\alpha)
\exp( 1/2 \cdot q_{\alpha\beta }c_\alpha^\dagger c_\beta^\dagger) 
\ket0 ,~~
r_\alpha
\!=\!
{\displaystyle \frac{1}{1 \!+\! z}}
(x_\alpha \!+\! q_{\alpha \beta } \overline{x}_\beta) ,
\label{SO2Nplus1wf}
\eeqa
\vspace{-0.8cm}
\beqa
\bra0\, U(G)\,\ket0 
\!=\! 
\overline{\Phi }_{00}(G) 
\!=\! 
\sqrt{\frac{1 \!+\! z}{2}}
\left[
\det(1_N \!+\! q^\dag q)
\right]^{1 / 4}
e^{i \tau / 2} .
\label{SO2Nplus1vacuumf}
\eeqa
The $SO(2N \!+\! 1)$ group is embedded into an $SO(2N \!+\! 2)$ group.
The embedding leads us to an unified formulation of the $SO(2N \!+\! 1)$
regular representation in which paired and unpaired modes are
treated in an equal way.
Define 
$(N \!+\! 1) \!\times\! (N \!+\! 1)$ matrices ${\cal A}$ and ${\cal B}$ as
\beq
{\cal A}
\!=\!
\left[ \!
\BA{cc}
A & -\overline{x} / 2 \\
\\
y / 2 & (1+z)/2
\EA \!
\right],
~~~
{\cal B}
\!=\!
\left[ \!
\BA{cc}
B & x / 2 \\
\\
-y / 2 & (1-z)/2
\EA \!
\right],~~~
y \!=\! x^{\mbox{\scriptsize T}}a \!-\! x^{\dag }b .
\label{calAcalB}
\eeq
Imposing the ortho-normalization of the $G$,
matrices ${\cal A}$ and ${\cal B}$
satisfy the ortho-normalization condition and then form an $SO(2N \!+\! 2)$
matrix ${\cal G}$ represented as
\cite{Fuk.77}
\beq
{\cal G}
\!=\!
\left[ \!
\BA{cc}
{\cal A} & \overline{\cal B} \\
{\cal B} & \overline{\cal A}
\EA \!
\right],
~~~~
{\cal G}^{\dag } {\cal G}
\!=\!
{\cal G}{\cal G}^{\dag }
\!=\! 1_{2N \!+\! 2} ,~~
\det {\cal G} \!=\! 1 .
\label{calG}
\eeq
By using
(\ref{relAtoaXY}),
the matrices ${\cal A}$ and ${\cal B}$
can be decomposed as\\[-10pt]
\beq
{\cal A}
\!=\!
\left[ \!
\BA{cc}
1_N \!-\! \overline{x} r^{\mbox{\scriptsize T}} / 2 &
-\overline{x} / 2 \\
\\
(1 \!+\! z)r^{\mbox{\scriptsize T}} / 2 &
(1 \!+\! z)/2
\EA \!
\right] \!\!
\left[ \!
\BA{cc}
a & 0 \\
\\
0 & 1
\EA \!
\right],
~
{\cal B}
\!=\!
\left[ \!
\BA{cc}
1_N \!+\! x r^{\mbox{\scriptsize T}}q^{-1} / 2 &
x / 2 \\
\\
- (1 \!+\! z)r^{\mbox{\scriptsize T}}q^{-1} / 2 &
(1 \!-\! z)/2
\EA \!
\right] \!\!
\left[ \!
\BA{cc}
b & 0 \\
\\
0 & 1
\EA \!
\right] ,
\label{calApcalBp}
\eeq
from which we get
the inverse of ${\cal A},~{\cal A}^{-1}$ and
$SO(2N \!+\! 2) / U(N \!+\! 1)$ coset variable
${\cal Q}$
with the $N \!+\! 1$-th component as\\[-10pt]
\beq
{\cal A}^{-1}
\!=\!
\left[ \!
\BA{cc}
a^{-1} & 0 \\
\\
0 & 1
\EA \!
\right]
\left[ \!
\BA{cc}
1_N & \overline{x}/ (1 \!+\! z) \\
\\
- r^{\mbox{\scriptsize T}} & 1
\EA \!
\right] ,~~
{\cal Q}
\!=\!
{\cal B}{\cal A}^{-1}
\!=\!
\left[ \!
\BA{cc}
q & r \\
\\
-r^{\mbox{\scriptsize T}} & 0
\EA \!
\right]
\!=\!
-{\cal Q}^{\mbox{\scriptsize T}} .
\label{cosetvariable2}
\eeq
We denote the $(N \!+\! 1)$-dimension of the matrix ${\cal Q}$ by
the index 0 and use the indices $i,~j,~\cdots$.


\newpage

\setcounter{equation}{0}
\renewcommand{\theequation}{\arabic{section}.\arabic{equation}}

\section{Recapitulation of $\frac{SO(\!2N\!+\!2\!)}{U(\!N\!+\!1\!)}$ supersymmetric $\sigma$-model and Killing potential}

~~~Following I
we recapitulate a SUSY $\sigma$-model
on the coset space $\frac{SO(2N \!+\! 2)}{U(N \!+\! 1)}$
and its Lagrangian on the K\"{a}hler manifold,
the symmetric space $\frac{SO(2N \!+\! 2)}{U(N \!+\! 1)}$.
The simplest representation of ${\cal N} \!\!=\!\! 1$ SUSY
is a scalar multiplet 
$\phi \!=\! \{{\cal Q}, \psi_L, H\}$
where ${\cal Q}$ and $H$ are complex scalars and
$
\psi_L
\!\equiv\!
{\displaystyle \frac{1 \!+\! \gamma_5}{2}} \psi
$ 
is a left-handed chiral spinor.
The SUSY $\sigma$-model
can be constructed from the
$[N] \{=N(N \!+\! 1) / 2\}$
scalar multiplets
$\phi^{[\alpha]} 
\!\!=\!\!
\{{\cal Q}^{[\alpha]},\psi^{[\alpha]}_L,H^{[\alpha]}\}
({[\alpha]} \!=\! 1,\cdots,[N])$.
Let the K\"{a}hler manifold be 
the $\frac{SO(2N \!+\! 2)}{U(N \!+\! 1)}$ coset manifold
and
redenote the complex scalar fields ${\cal Q}_{pq}$ 
as 
${\cal Q}^{[\alpha]}([\alpha] \!=\! 1, \!\cdots\!, [N])$.
After eliminating the auxiliary field $H^{\![\alpha]}$,
the Lagrangian of a SUSY $\sigma$-model
\cite{NNH.01}
is given as\\[-16pt]
\beqa
\BA{c}
{\cal L}_{{\mbox{\scriptsize chiral}}}
\!=\!
-
{\cal G}_{[\alpha][\underline{\beta }]}
\left( \!
\partial_\mu \bar{{\cal Q}}^{[\underline{\beta }]}
\partial_\mu {\cal Q}^{[\alpha]}
\!+\!
\bar{\psi }^{[\underline{\beta }]}_L
\overleftrightarrow{ \mbox{\boldmath $\slashed{D}$} }
\psi^{[\alpha]}_L \!
\right)
\!+\!
{\displaystyle \frac{1}{2}}
\mbox{\boldmath $R$}
_{[\alpha][\underline{\beta }][\gamma][\underline{\delta }]}
\bar{\psi }^{[\underline{\beta }]}_L \gamma_\mu \psi^{[\alpha ]}_L
\bar{\psi }^{[\underline{\delta }]}_L \gamma_\mu \psi^{[\gamma ]}_L .
\EA
\label{susylaglangian}
\eeqa\\[-16pt]
The K\"{a}hler metrics admit a set of holomorphic isometries,
the Killing vectors,
${\cal R}^{l[\alpha]}({\cal Q})$
and
$\overline{\cal R}^{l[\underline{\alpha }]}(\overline{\cal Q})
(l \!=\! 1, \cdots, \dim {\cal G}({\cal G} \!\in\! SO(2N\!+\!2)))$,
which are the solution of the Killing equation\\[-10pt]
\beq
{\cal R}^l _{~[\underline{\beta }]}({\cal Q})_{,[\alpha]}
\!+\!
\overline{\cal R}^l _{~[\alpha]}(\overline{\cal Q})_{,[\underline{\beta }]}
\!=\!
0 ,~~
{\cal R}^l _{~[\underline{\beta }]}({\cal Q})
\!=\!
{\cal G}_{[\alpha][\underline{[\beta }]}{\cal R}^{l[\alpha]}({\cal Q}) .
\label{Killingeq}
\eeq
These isometries
are described geometrically by the above Killing vectors,
the generators of infinitesimal coordinate transformations
keeping the metric invariant: 
$
\delta {\cal Q}
\!=\!
{\cal Q}^\prime \!-\! {\cal Q}
\!=\!
{\cal R}({\cal Q})
$
and
$
\delta \overline{\cal Q}
\!=\!
\overline{\cal R}(\overline{\cal Q})
$
such that
$
{\cal G}^\prime ({\cal Q}, \overline{\cal Q})
\!=\!
{\cal G} ({\cal Q}, \overline{\cal Q})
$.
The Killing equation
(\ref{Killingeq})
is the necessary and sufficient condition for 
an infinitesimal coordinate transformation\\[-10pt]
\beq
\delta{\cal Q}^{[\alpha]}
\!=\!
\left(
\delta {\cal B} 
- \delta {\cal A}^{\mbox{\scriptsize T}}{\cal Q} \!-\! {\cal Q}\delta {\cal A}
\!+\! {\cal Q}\delta {\cal B}^\dag {\cal Q}
\right)^{[\alpha]}
\!=\!
\widehat{\xi }_l{\cal R}^{l[\alpha]}({\cal Q}) ,~~
\delta \overline{\cal Q}^{[\underline{\alpha }]}
\!=\!
\widehat{\xi }_l \overline{\cal R}^{l[\underline{\alpha }]}
(\overline{\cal Q}) ,
\label{infinitesimaltrans}
\eeq
where
$\widehat{\xi }_l$
are infinitesimal parameters.
Due to the Killing equation,
the Killing vectors
${\cal R}^{l[\alpha]}({\cal Q})$
and
$\overline{\cal R}^{l[\underline{\alpha }]}(\overline{\cal Q})$
are given as the gradient of
the Killing potential
${\cal M}^l ({\cal Q}, \overline{\cal Q})$
such that\\[-12pt]
\beq
{\cal R}^l _{~[\underline{\alpha }]}({\cal Q})
\!=\!
-i{\cal M}^l _{~,[\underline{\alpha }]} ,~~
\overline{\cal R}^l _{~[\alpha]}(\overline{{\cal Q}})
\!=\!
i{\cal M}^l _{~,[\alpha]} .
\label{gradKillingpot}
\eeq
The Killing potential ${\cal M}_\sigma$ 
can be written as\\[-20pt]
\beqa
\BA{c}
{\cal M}_\sigma \!
\left(
\delta {\cal A}, \delta {\cal B},\delta {\cal B}^\dag
\right)
\!=\!
\mbox{Tr} \!
\left(
\delta {\cal G} \widetilde{{\cal M}}_\sigma
\right)
\!=\!
\mbox{tr} \!
\left(
\delta {\cal A} {\cal M}_{\sigma \delta {\cal A}}
\!+\!
\delta {\cal B} {\cal M}_{\sigma \delta {\cal B}^\dag }
\!+\!
\delta {\cal B}^\dag {\cal M}_{\sigma \delta {\cal B}}
\right) .
\EA
\label{KillingpotM}
\eeqa\\[-18pt]
Trace Tr extends over 
the $(2N \!+\! 2) \!\times\! (2N \!+\! 2)$ matrices,
while trace tr does over 
the $(N \!+\! 1) \!\times\! (N \!+\! 1)$ matrices.
Let us introduce the $(N \!+\! 1)$-dimensional matrices 
${\cal R}({\cal Q}; \delta {\cal G})$, 
${\cal R}_T({\cal Q}; \delta {\cal G})$ and ${\cal X}$ by\\[-16pt]
\beqa
\left.
\BA{ll}
&
{\cal R}({\cal Q}; \delta {\cal G})
\!=\!
\delta {\cal B} 
\!-\! \delta {\cal A}^{\mbox{\scriptsize T}}{\cal Q}
\!-\! {\cal Q}\delta {\cal A}
\!+\!
{\cal Q}\delta {\cal B}^\dag {\cal Q} ,~~
{\cal R}_T ({\cal Q}; \delta {\cal G})
\!=\!
-\delta {\cal A}^{\mbox{\scriptsize T}}
\!+\!
{\cal Q}\delta {\cal B}^\dag ,\\
\\[-8pt]
&
{\cal X}
\!=\!
(1_{N \!+\! 1} \!+\! {\cal Q}{\cal Q}^\dag)^{-1}
\!=\!
{\mathcal X}^\dag .
\EA
\right\}
\label{RRTChi}
\eeqa\\[-12pt]
In 
(\ref{infinitesimaltrans}),
$\delta {\cal Q}
\!=\! 
{\cal R}({\cal Q}; \delta {\cal G})
$,
Killing vector
in the coset space
$\frac{SO(2N \!+\! 2)}{U(N \!+\! 1)}$
and
$
\mbox{tr} [{\cal R}_T ({\cal Q}; \delta {\cal G})]
\!=\! {\cal F}({\cal Q})
$
is derived later
where
${\cal F}({\cal Q})$
is a holomorphic function.
The Killing potential ${\cal M}_\sigma$ is given as\\[-16pt]
\beqa
\!\!\!\!\!\!\!\!
\left.
\BA{rl}
&-i{\cal M}_\sigma \!
\left(
{\cal Q}, \bar{\cal Q};\delta {\cal G}
\right)
\!=\!
-\mbox{tr}
\Delta \!
\left(
{\cal Q}, \bar{\cal Q};\delta {\cal G}
\right) ,\\
\\[-8pt]
&\Delta \!
\left(
{\cal Q}, \bar{\cal Q};\delta {\cal G}
\right)
\stackrel{\mathrm{def}}{=}
{\cal R}_T ({\cal Q}; \delta {\cal G})
\!-\!
{\cal R}({\cal Q}; \delta {\cal G}) {\cal Q}^\dag {\cal X} 
\!=\!
\left(
{\cal Q} \delta {\cal A} {\cal Q}^\dag 
\!-\!
\delta {\cal A}^{\mbox{\scriptsize T}}
\!-\!
\delta {\cal B} {\cal Q}^\dag 
\!+\!
{\cal Q} \delta {\cal B}^\dag
\right) \!
{\cal X} .
\EA \!\!
\right\}
\label{formKillingpotM}
\eeqa\\[-12pt]
From
(\ref{KillingpotM}) and (\ref{formKillingpotM}),
we obtain\\[-14pt]
\beq
-i{\cal M}_{\sigma \delta {\cal B}}
\!=\!
-{\cal X} {\cal Q} ,~~
-i{\cal M}_{\sigma \delta {\cal B}^\dag }
\!=\!
{\cal Q}^\dag {\cal X} ,~~
-i{\cal M}_{\sigma \delta {\cal A}}
\!=\!
1_{N+1} \!-\! 2 {\cal Q}^\dag {\cal X} {\cal Q} .
\label{componentKillingpotM} 
\eeq\\[-20pt]
To make clear the meaning of the Killing potential,
using the $(2N \!+\! 2) \!\times\! (N \!+\! 1)$ isometric matrix 
${\cal U}~
({\cal U}^{\mbox{\scriptsize T}}
\!=\!
\left[ \!
\BA{cc} 
{\cal B}^{\mbox{\scriptsize T}}, {\cal A}^{\mbox{\scriptsize T}}
\EA \!
\right] ,
{\cal U}^\dag{\cal U}\!=\!1_{N \!+\! 1})
$,
let us define the following 
$(2N \!+\! 2) \!\times\! (2N \!+\! 2)$ matrix:\\[-6pt]
\beq
{\cal W}
\stackrel{\mathrm{def}}{=}
{\cal U}{\cal U}^\dag
\!=\!
\left[ 
\BA{cc} 
{\cal R} & {\cal K} \\
\\[-8pt]
-\overline{\cal K} & 1_{N \!+\! 1} - \overline{\cal R}
\EA 
\right] 
\!=\!
{\cal W}^\dag ~
({\cal W}^2 \!=\! {\cal W}) ,
\left\{ \!
\BA{c}
{\cal R}
\!=\!
{\cal B}{\cal B}^\dag ,\\
\\[-8pt]
{\cal K}
\!=\!
{\cal B}{\cal A}^\dag .
\EA
\right.
\label{densitymat}
\eeq
The ${\cal W}$ is a natural extension of 
the generalized density matrix in the $SO(2N)$ CS rep
to the $SO(2N \!+\! 2)$ CS rep.
The matrices ${\cal R}$ and ${\cal K}$
are represented in terms of ${\cal Q}$ and ${\cal X}$ as\\[-10pt]
\beq
{\cal R}
\!=\!
{\cal Q}(1_{N \!+\! 1} \!+\! {\cal Q}^\dag{\cal Q})^{-1}{\cal Q}^\dag
\!=\!
{\cal Q}\bar{{\cal X}}{\cal Q}^\dag
\!=\!
1_{N \!+\! 1} \!-\! {\cal X} ,~~
{\cal K}
\!=\!
{\cal Q}
(1_{N \!+\! 1} \!+\! {\cal Q}^\dag{\cal Q})^{-1}
\!=\!
{\cal X} {\cal Q} .
\label{matRandK}
\eeq\\[-14pt]
Using
(\ref{matRandK})
and
(\ref{componentKillingpotM}),
the $-i\overline{\widetilde{{\cal M}}}_\sigma$
reads
the generalized density matrix
(\ref{densitymat}).

Introducing the gauge fields in Lagrangian
(\ref{susylaglangian}),
via the gauge covariant derivatives,
the $\sigma$-model is no longer invariant under the SUSY transformations.
To restore the SUSY,
it is necessary to add the terms\\[-22pt]
\beq
\Delta {\cal L}_{\mbox{{\scriptsize chiral}}}
\!=\!
2 {\cal G}_{[\alpha][\underline{\alpha }]} \!\!
\left( \!\!
{\cal R}^l _{~[\underline{\alpha }]}({\cal Q})
\bar{\psi }^{[\underline{\alpha }]}_L \lambda^l _R
\!+\!
\overline{\cal R}^l _{~[\alpha]}(\overline{\cal Q})
\overline{\lambda }^l _R \psi^{[\alpha]}_L \!\!
\right)
\!-\!
g_l \mbox{tr} \!
\left\{ \! D^l ({\cal M}^l \!+\! \xi^l) \! \right\} \! ,~
\lambda^l _R
\!=\!
\frac{1 \!-\! \gamma^5}{2}
\lambda^l ,
\label{addingterm}
\eeq\\[-16pt]
where $\xi_l$ are Fayet-Ilipoulos parameters.
Then the full Lagrangian for this model consists of
the usual SUSY Yang-Mills part and the chiral part\\[-6pt]
\beq
{\cal L}
\!=\!
-
\mbox{tr}
\left\{
\frac{1}{4}{\cal F}^l _{\mu \nu }{\cal F}^l _{\mu \nu }
\!+\!
\frac{1}{2}\overline{\lambda }^l \slashed{D} \lambda^l
\!-\!
\frac{1}{2}D^l D^l
\right\}
\!+\!
{\cal L}_{{\mbox{\scriptsize chiral}}}
(\partial_\mu \!\rightarrow\! \mbox{\boldmath $D$}_\mu)
\!+\!
\Delta {\cal L}_{\mbox{{\scriptsize chiral}}} .
\label{fullLaglangian}
\eeq\\[-10pt]
Eliminating the auxiliary field $D^l$ by 
$
D^l
\!=\!
-g_l ({\cal M}^l \!+\! \xi^l)
$
(not summed for $l$),
we get a RESP arising from the gauging of
$SU(N \!+\! 1) \!\times\! U(1)$ with
a Fayet-Ilipoulos term $\xi$\\[-16pt]
\beqa
\left.
\BA{rl}
&\!\!\!\!
V_{{\mbox{\scriptsize redSC}}}
\!=\!
{\displaystyle \frac{g^2 _{U(1)}}{2(N \!+\! 1)}}
\left(
\xi - i {\cal M}_Y
\right)^2
\!+\!
{\displaystyle \frac{g^2 _{SU(N+1)}}{2}}
\mbox{tr} \!
\left(
- i {\cal M}_t
\right)^2 , \\
\\[-12pt]
&\!\!\!\!
\mbox{tr} \!
\left( - i {\cal M}_t \right)^2
\!=\!
\mbox{tr} \!
\left( - i {\cal M}_{\sigma \delta {\cal A}} \right)^2
\!-\!
{\displaystyle \frac{1}{N+1}}
(- i {\cal M}_Y)^2 ,~
- i {\cal M}_Y
\!=\!
\mbox{tr} \!
\left(
- i {\cal M}_{\sigma \delta {\cal A}} 
\right) .
\EA
\right\}
\label{specialscalarpot}
\eeqa

To find an $f$-deformed solution
of the SUSY $\sigma\!$-model,
we introduce the $(N \!+\! 1)$-dimensional matrices
${\cal R}_f ({\cal Q}_f; \delta {\cal G})$,
${\cal R}_{fT}({\cal Q}_f; \delta {\cal G})$ and ${\cal X}_f$
in the following forms:\\[-16pt]
\beqa
\!\!\!\!\!\!\!\!\!\!\!\!
\left.
\BA{ll}
&
{\cal R}_f ({\cal Q}_f; \delta {\cal G})
\!=\!
{\displaystyle \frac{1}{f}}
\delta {\cal B} 
\!-\! 
\delta {\cal A}^{\mbox{\scriptsize T}}{\cal Q}_f 
\!-\! 
{\cal Q}_f \delta {\cal A}
\!+\! 
f {\cal Q}_f \delta {\cal B}^\dag {\cal Q}_f ,~
{\cal R}_{fT} ({\cal Q}_f; \delta {\cal G})
\!=\!
\!-\!
\delta {\cal A}^{\mbox{\scriptsize T}}
\!+\! 
f {\cal Q}_f \delta {\cal B}^\dag , \!\! \\
\\[-12pt]
&
{\cal X}_f
\!=\!
(1_{N+1} + f^2 {\cal Q}_f {\cal Q}^\dag _f)^{-1}
\!=\! 
{\mathcal X}^\dag ,~
{\cal Q}_f
\!=\! 
\left[ \!\!
\BA{cc}
q & {\displaystyle \frac{1}{f}} r_{\!f} \\
- {\displaystyle \frac{1}{f}} r^{\mbox{\scriptsize T}} _{\!f} & 0
\EA \!\!
\right] ,~
r_{\!f}
\!=\!
{\displaystyle \frac{1}{2 Z^2}}
\left(
x \!+\! f q \overline{x}
\right) ,~
{\displaystyle f 
\!\stackrel{\mathrm{def}}{=}\! 
\frac{1}{m_\sigma }} . \!\!
\EA
\right\}
\label{RRTChif}
\eeqa
Due to the rescaling, 
the Killing potential ${\cal M}_{\sigma }$ is deformed as
\beqa
\left.
\BA{rl}
-i{\cal M}_{f \sigma } \!
\left(
{\cal Q}_f, \bar{\cal Q}_f;\delta {\cal G}
\right)
\!\!&\!=\!
-\mbox{tr}
\Delta_f \!
\left(
{\cal Q}_f, \bar{\cal Q}_f;\delta {\cal G}
\right) ,\\
\\[-8pt]
\Delta_f \!
\left(
{\cal Q}_f, \bar{\cal Q}_f;\delta {\cal G}
\right)
\!\!&\!
\stackrel{\mathrm{def}}{=}\!
{\cal R}_{fT} ({\cal Q}_f; \delta {\cal G})
\!-\!
{\cal R}_f({\cal Q}_f; \delta {\cal G}) f^2 {\cal Q}^\dag _f {\cal X}_f \\
\\[-6pt]
\!\!&
\!=\!
\left( \!
f^2 {\cal Q}_f \delta {\cal A} {\cal Q}^\dag _f
\!-\!
\delta {\cal A}^{\mbox{\scriptsize T}}
\!-\!
f \delta {\cal B} {\cal Q}^\dag _f
\!+\!
f {\cal Q}_f \delta {\cal B}^\dag \!
\right) \!
{\cal X}_f ,
\EA
\right\}
\label{formKillingpotMf}
\eeqa
from which we obtain
a $f$-deformed Killing potential ${\cal M}_{f \sigma }$\\[-12pt]
\beq
-i{\cal M}_{f \sigma \delta {\cal B}}
\!=\!
-f {\cal X}_f {\cal Q}_f ,~
-i{\cal M}_{f \sigma \delta {\cal B}^\dag }
\!=\!
 f {\cal Q}^\dag _f {\cal X}_f ,~
-i{\cal M}_{f \sigma \delta {\cal A}}
\!=\!
1_{N+1} \!-\! 2 f^2 {\cal Q}^\dag _f{\cal X}_f {\cal Q}_f .
\label{componentKillingpotMf}
\eeq
After the same algebraic manipulations, 
the inverse matrix 
${\cal X}_f$
in (\ref{RRTChif}) 
leads to the form
\beq
{\mathcal X}_f
\!=\!
\left[ \!\!\!\!
\BA{cc}
{\cal Q}_{f qq^\dag } &\!\! {\cal Q}_{f q \bar{r}} \\
\\[-6pt]
{\cal Q}_{f q \bar{r}}^{\dag } &\!\! {\cal Q}_{f r^\dag r}
\EA \!\!\!\!
\right] \! ,~
\chi_{\!f}
\!=\!
(1_N \!+\! f^2 q q^\dag )^{-1} 
\!=\!
\chi^\dag _{\!f} ,
\BA{c}
{\cal Q}_{f qq^\dag }
\!=\!
\chi_{\!f} \!-\! Z^2 \chi_{\!f} (r_{\!f} r^\dag _f
\!-\! f^2 q\bar{r}_{\!f} r^{\mbox{\scriptsize T}}_{\!f} q^\dag )
\chi_{\!f} ,\\
\\[-8pt]
{\cal Q}_{f q \bar{r}}
\!=\!
f Z^2 \chi_f q \bar{r}_f ,~~~~~
{\cal Q}_{f r^\dag r}
\!=\!
Z^2 .
\EA
\label{inverse1plusQQf}
\eeq
Substituting 
(\ref{RRTChif}) and (\ref{inverse1plusQQf})
into
(\ref{componentKillingpotMf})
and introducing a $f$-deformed auxiliary function
$
\lambda_f
\!=\!
r_{\!f} r_{\!f} ^\dag  \!-\! f^2 q\overline{r}_{\!f}
r_f ^{\mbox{\scriptsize T}}q^\dag
\!=\!
\lambda^\dag _{\!f}
$,
we can get the $f$-deformed Killing potential
${\cal M}_{f \sigma \delta {\cal A}}$
as
\vspace{-0.1cm}
\beqa
\!\!\!\!\!\!\!\!
\BA{ll}
&~
-i{\cal M}_{f \sigma \delta {\cal A}} 
= \\
\\[-10pt]
&
\left[ \!\!\!\!\! 
\BA{cc}
\BA{c}
1_N \!-\! 2q^\dag \chi_{\!f} q 
\!+\! 
2
{\displaystyle \frac{Z^2}{f^2}} \!
\left( \!
f^2 q^\dag \chi_{\!f} \lambda_{\!f} \chi_{\!f} q
\!+\!
f^2 q^\dag \chi_{\!f} q \overline{r}_{\!f}
r^{\mbox{\scriptsize T}}_{\!f}
\right. \\
\left.
+
f^2 \overline{r}_{\!f} r^{\mbox{\scriptsize T}}_{\!f}
q^\dag \chi_{\!f} q
\!-\!
\overline{r}_{\!f} r^{\mbox{\scriptsize T}}_{\!f}
\right)
\EA
&
\BA{c} 
-
2 {\displaystyle \frac{1}{f}}q^\dag \chi_{\!f} r_{\!f}
\!+\! 
2 {\displaystyle \frac{Z^2}{f}} \!
\left(
q^\dag \chi_{\!f} \lambda_{\!f} \chi_{\!f} r_{\!f}
\right. \\
\left.
+
\overline{r}_{\!f} r^{\mbox{\scriptsize T}}_{\!f}
q^\dag \chi_{\!f} r_{\!f}
\right)
\EA \\
\\[-12pt]
\BA{c}
- 2 {\displaystyle \frac{1}{f}} r^\dag_{\!f} \chi_{\!f} q
\!+\! 
2 {\displaystyle \frac{Z^2}{f}} \!
\left( \!
r^\dag_{\!f} \chi_{\!f} \lambda_{\!f} \chi_{\!f} q
+
r^\dag_{\!f} \chi_{\!f} q \overline{r}_{\!f}
r^{\mbox{\scriptsize T}}_{\!f}
\right)
\EA
&\!\!\!\!
1
\!-\!
2 {\displaystyle \frac{1}{f^2}}r^\dag_{\!f} \chi_{\!f} r_{\!f}
\!+\!
2{\displaystyle \frac{Z^2}{f^2}}
r^\dag_{\!f} \chi_{\!f} \lambda_{\!f} \chi_{\!f} r_{\!f}
\EA \!\!\!\!\!
\right] \! .
\EA
\label{KillingpotAf}
\eeqa\\[-8pt]
Using
$
r_f
\!=\!
{\displaystyle
\frac{1}{2Z^2}
(x \!+\! f q \overline{x})
}
$
again,
the following relations also can be easily proved:\\[-8pt]
\beq
1
\!-\!
2 \frac{1}{f^2} r^\dag_{\!f} \chi_{\!f} r_{\!f}
\!+\!
2 \frac{Z^2}{f^2}  
r^\dag_{\!f} \chi_{\!f} \lambda_{\!f} \chi_{\!f} r_{\!f}
\!=\!
\frac{1}{f^2} (2Z^2 \!-\! 1) \!+\! 1 \!-\! \frac{1}{f^2} ,
\label{relation1f}
\eeq
\vspace{-0.3cm}
\beq
\chi_{\!f} \lambda_{\!f} \chi_{\!f} r_{\!f}
\!=\!
\frac{1 \!-\! Z^2}{Z^2} \chi_{\!f} r_{\!f} ,~~
r^\dag_f \chi_f \lambda_f \chi_f
\!=\!
\frac{1 \!-\! Z^2}{Z^2} r^\dag_{\!f} \chi_{\!f} ,~~ 
q^\dag \chi_{\!f} q
\!=\!
\frac{1}{f^2} (1_N \!-\! \bar{\chi }_f) ,
\label{relation2f}
\eeq
from which,
we get a more
compact form of the $f$-deformed Killing potential
${\cal M}_{f \sigma \delta {\cal A}}$
as,
\vspace{-0.1cm}
\beq
\!\!
-i{\cal M}_{f \sigma \delta {\cal A}}
\!=\!
\left[ \!\!\!
\BA{cc}
1_N \!-\! 2q^\dag \chi_{\!f} q
\!+\!
2
{\displaystyle \frac{Z^2}{f}} \!
\left( \!\!
f q^\dag \chi_{\!f} r_{\!f} r^\dag_{\!f} \chi_{\!f} q
\!-\!
{\displaystyle \frac{1}{f}}
\overline{\chi }_{\!f} \overline{r}_{\!f}
r^{\mbox{\scriptsize T}}_{\!f} \overline{\chi }_{\!f} \!\!
\right)  &
- 2
{\displaystyle \frac{Z^2}{f}} q^\dag \chi_{\!f} r_{\!f} \\
\\[-12pt]
- 2 {\displaystyle \frac{Z^2}{f}}
r^\dag_{\!f} \chi_{\!f} q & 
{\displaystyle \frac{1}{f^2}
(2Z^2 \!-\! 1) \!+\! 1 \!-\! \frac{1}{f^2}}
\EA \!\!\!
\right] .
\label{KillingpotA2f}
\eeq
Owing to the rescaling,
the $f$-deformed reduced scalar potential is written
as follows:\\[-20pt]
\beqa
\left.
\BA{rl}
&\!\!\!\!
V_{\!f{\mbox{\scriptsize redSC}}}
\!=\!
{\displaystyle \frac{g^2 _{U(1)}}{2(N \!+\! 1)}}
\left(
\xi - i {\cal M}_{fY}
\right)^2
\!+\!
{\displaystyle \frac{g^2 _{SU(N+1)}}{2}}
\mbox{tr} \!
\left(
- i {\cal M}_{ft}
\right)^2 , \\
\\[-12pt]
&\!\!\!\!
\mbox{tr} \!
\left( - i {\cal M}_{ft} \right)^2
\!=\!
\mbox{tr} \!
\left( - i {\cal M}_{f \sigma \delta {\cal A}} \right)^2
\!-\!
{\displaystyle \frac{1}{N \!+\! 1}} \!
\left( - i {\cal M}_{fY} \right)^2 ,~
- i {\cal M}_{fY}
\!=\!
\mbox{tr} \!
\left(
- i {\cal M}_{f \sigma \delta {\cal A}}
\right),
\EA
\right\}
\label{specialscalerpotf}
\eeqa
in which each $f$-deformed Killing potential is computed
straightforwardly as\\[-16pt]
\beqa
\!\!\!\!
\mbox{tr} \!
\left( \!
-i{\cal M}_{f \sigma \delta {\cal A}} \!
\right)
\!\!=\!\!
\left( \!\!
1 \!\!-\!\! 2 {\displaystyle \frac{1}{f^2}} \!\!
\right) \!\!
N
\!\!+\!\!
2 {\displaystyle \frac{1}{f^2}} \! \mbox{{\bf tr}}(\chi_{\!f})
\!\!+\!\!
2 {\displaystyle \frac{Z^2}{f^2}} \!
\mbox{{\bf tr}}(\chi_{\!f} r_{\!f} r^\dag_{\!f})
\!\!-\!\!
4 {\displaystyle \frac{Z^2}{f^2}} \!
\mbox{{\bf tr}}(\chi_{\!f} r_{\!f} r^\dag_{\!f} \chi_{\!f})
\!\!+\!\!
{\displaystyle \frac{1}{f^2}} \!
( \! 2 Z^2 \!\!-\!\! 1 \! )
\!\!+\!\!
1 \!\!-\!\! {\displaystyle \frac{1}{f^2}} ,
\label{trM2f1}
\eeqa
\vspace{-0.4cm}
\def\thefootnote{\fnsymbol{footnote}}
\beqa
\!\!\!\!\!\!\!\!
\BA{rl}
&\!\!\!\!
\mbox{tr} \!
\left(
-i{\cal M}_{f \sigma \delta {\cal A}}
\right)^2
\!=\!
N \!\!-\! 4 {\displaystyle \frac{1}{f^2}}
(1 \!-\! {\displaystyle \frac{1}{f^2}}) N
\!\!-\!
4 {\displaystyle \frac{1}{f^4}} \mbox{{\bf tr}}(\chi_{\!f} )
\!+\!
4 {\displaystyle \frac{1}{f^4}}
\mbox{{\bf tr}}(\chi_{\!f} \chi_{\!f}) 
\!+\!
4 {\displaystyle \frac{1}{f^2}}
(1 \!-\! {\displaystyle \frac{1}{f^2}})
< \!\chi_{\!f}\! > \!\footnotemark[1]\\
\\[-12pt]
&
\!+\!
4 (1 \!-\! {\displaystyle \frac{1}{f^2}})
{\displaystyle \frac{Z^2}{f^2}}
\left\{
\mbox{{\bf tr}}(\chi_{\!f} r_{\!f} r^\dag_{\!f})
\!-\!
\mbox{{\bf tr}}(\chi_{\!f} r_{\!f} r^\dag_{\!f} \chi_{\!f})
\right\}
+
12 {\displaystyle \frac{Z^2}{f^4}}
\mbox{{\bf tr}}(\chi_{\!f} r_{\!f} r^\dag_{\!f} \chi_{\!f})
\!-\!
16 {\displaystyle \frac{Z^2}{f^4}}
\mbox{{\bf tr}}(\chi_{\!f} \chi_{\!f}
r_{\!f} r^\dag_{\!f} \chi_{\!f}) \\
\\[-12pt]
&
-
4 {\displaystyle \frac{Z^4}{f^4}} r^\dag_{\!f}
\chi_{\!f} \chi_{\!f} r_{\!f}
\cdot
\mbox{{\bf tr}}(\chi_{\!f} r_{\!f} r^\dag_{\!f})
\!+\!
8 {\displaystyle \frac{Z^4}{f^4}} r^\dag_{\!f}
\chi_{\!f} \chi_{\!f} r_{\!f}
\cdot
\mbox{{\bf tr}}(\chi_{\!f} r_{\!f} r^\dag_{\!f} \chi_{\!f}) \\
\\[-12pt]
&
+
{\displaystyle \frac{1}{f^4}}
\!+\!
2 {\displaystyle \frac{1}{f^2}}
(1 \!-\! {\displaystyle \frac{1}{f^2}}) (2 Z^2 \!-\! 1)
\!+\!
(1 \!-\! {\displaystyle \frac{1}{f^2}})^2
\!-\!
4 {\displaystyle \frac{Z^4}{f^4}} r^\dag_{\!f}
\chi_{\!f} \chi_{\!f} r_{\!f} .
\EA
\label{trM2f2}
\eeqa
\footnotetext[1]
{We take the opportunity of pointing out a misprint in Eq. (5.12) in I
\cite{SJCF.08},
where the last term in the first line of
(\ref{trM2f2})
is missing.}
The trace
$\mbox{{\bf tr}}$,
taken over the $N \!\times\! N$ matrix, is used.
The
$r^\dag_{\!f} \chi_{\!f} \chi_{\!f} r_{\!f}$ and
$\mbox{{\bf tr}}(r_{\!f} r^\dag_{\!f})$
are approximately computed
as
\beqa
\left.
\BA{ll}
r^\dag_{\!f} \chi_{\!f} \chi_{\!f} r_{\!f}
\!=\!&\!\!\!\!
{\displaystyle
\frac{1}{4Z^4}
x^\dag \chi_{\!f} x
} 
\!\approx\!
{\displaystyle \frac{1 \!-\! Z^2}{Z^2}} \!
< \!\chi_{\!f}\! >, ~
< \!\chi_{\!f}\! >
\stackrel{\mathrm{def}}{=}\!
\left\{ \!
{\displaystyle \frac{1}{N}} \!
\left[
N \!+\! f^2 \mbox{{\bf tr}}(q^\dag q)
\right] \!
\right\}^{\!-1} \!\!
\!=\!
\overline{< \!\chi_{\!f}\! >} , \\
\\[-12pt]
\mbox{{\bf tr}}(r_f r^\dag_f )
\!=\!&\!\!\!\!
r^\dag_{\!f} r_{\!f}
\!=\!
{\displaystyle
\frac{1}{4Z^4}
x^\dag \chi_{\!f} ^{-1} x
}
\!\approx
< \!r_{\!f} r^\dag_{\!f}\! >,~
< \!r_{\!f} r^\dag_{\!f}\! >
\stackrel{\mathrm{def}}{=}
{\displaystyle\!
\frac{1 \!-\! Z^2}{Z^2}
\frac{1}{< \!\chi_{\!f}\! >}
} .
\EA \!
\right\}
\label{rdagchichirftrrrdagf}
\eeqa
In 
(\ref{trM2f1}) and (\ref{trM2f2}),
approximating
$\mbox{{\bf tr}}(\chi_{\!f})$,
$\mbox{{\bf tr}}(\chi_{\!f} r_{\!f} r^\dag_{\!f})$, etc.
by
$< \!\!\chi_{\!f}\!\! >$,
$< \!\!\chi_{\!f}\!\! > \!
\mbox{{\bf tr}}(r_{\!f}r^\dag_{\!f})$, etc.,
respectively,
and using
(\ref{rdagchichirftrrrdagf}),
$
\mbox{tr} \!
\left(
-i{\cal M}_{f \sigma \delta {\cal A}}
\right)
$
and 
$
\mbox{tr} \!
\left(
-i{\cal M}_{f \sigma \delta {\cal A}}
\right)^2
$
are computed as\\[-18pt]
\beqa
\!\!\!\!\!\!
\left.
\BA{rl}
\mbox{tr} \!
\left(
-i{\cal M}_{f \sigma \delta {\cal A}}
\right)
\!=\!
&\!\!\!\!
1 
\!+\!
\left( \!
1 \!-\! 2 {\displaystyle \frac{1}{f^2}} \!
\right) \!
N
\!+\!
2 {\displaystyle \frac{1}{f^2}} (2Z^2 \!-\! 1) \!
< \!\chi_{\!f}\! > , \\
\\[-8pt]
\mbox{tr} \!
\left(
-i{\cal M}_{f \sigma \delta {\cal A}}
\right)^2
\!=\!
&\!\!\!\!
1 \!+\! N \!-\! 4 {\displaystyle \frac{1}{f^2}}
\left( \!
1 \!-\! {\displaystyle \frac{1}{f^2}} \!
\right) \! N \\
\\[-14pt]
\!-\!
&\!\!\!\!
4 {\displaystyle \frac{1}{f^2}} \!
\left\{ \!
{\displaystyle \frac{1}{f^2}}(2 Z^2 \!-\! 1)
\!-\!
\left( \!
1 \!-\! {\displaystyle \frac{1}{f^2}} \!
\right) \!
Z^2 \!
\right\} \!
< \!\chi_{\!f}\! >
\!+\! 
4 {\displaystyle \frac{1}{f^4}}(2 Z^4 \!-\! 1) \!
< \!\chi_{\!f}\! >^2 .
\EA \!\!
\right\}
\label{approxKillingpotf}
\eeqa\\[-10pt]
Substituting
(\ref{approxKillingpotf})
into
(\ref{specialscalerpotf}),
we obtain the $f$-deformed reduced scalar potential
as\\[-18pt]
\beqa
\!\!\!\!\!\!
\BA{rl}
&\!\!\!\!
V_{\!f{\mbox{\scriptsize redSC}}}
\!=\!
{\displaystyle
\frac{g^2 _{U(1)}}{2(N \!+\! 1)}
}
\left[
\xi \!+\! 1
\!+\!
\left( \! 1 \!-\! 2 {\displaystyle \frac{1}{f^2}} \! \right) \!
N 
\!+\!
2
{\displaystyle \frac{1}{f^2}}(2 Z^2 \!-\! 1)  
< \!\chi_{\!f}\! >
\right]^2 \\
\\[-10pt]
&\!\!\!\!
\!+\!
{\displaystyle
2 
\frac{g^2 _{SU(N \!+\! 1)}}{N \!+\! 1}
}
{\displaystyle \frac{1}{f^2}}
\left[
{\displaystyle \frac{1}{f^2}} N
\!-\!
\left\{ \!
\left( \!
1 \!-\! {\displaystyle \frac{1}{f^2}} \!
\right) \! N
\!+\!
\left( \!
1 \!+\! 3 {\displaystyle \frac{1}{f^2}} \!
\right) \!
\right\} \!
Z^2 \! < \!\chi_{\!f}\! >
\right. \\
\\[-10pt]
&\!\!\!\!
\left.
\!+\!
\left\{ \!
\left( \!
1 \!-\! {\displaystyle \frac{1}{f^2}} \!
\right) N
\!+\!
\left( \!
1 \!+\! {\displaystyle \frac{1}{f^2}} \!
\right) \!
\right\}
< \!\chi_{\!f}\! >
\!+\!
{\displaystyle \frac{1}{f^2}}
\left\{
2(N \!-\! 1) Z^4 \!+\! 4 Z^2 \!-\! (N \!+\! 2)
\right\} \!
< \!\chi_{\!f}\! >^2
\right] .
\EA
\label{specialscalerpot2f}
\eeqa\\[-10pt]
The variation of
(\ref{specialscalerpot2f})
with respect to $Z^2$ and $< \!\!\chi_{\!f}\!\! >$
reads\\[-20pt]
\beqa
\BA{rl}
&\!\!\!\!
g^2 _{U(1)}
\left\{ \!
\xi \!+\! 1
\!+\!
\left( \! 1 \!-\! 2 {\displaystyle \frac{1}{f^2}} \! \right) \!
N
\!+\!
2
{\displaystyle \frac{1}{f^2}}(2 Z^2 \!-\! 1)
< \!\chi_{\!f}\! > \!
\right\} \\
\\[-10pt]
&\!\!\!\!
-
2 
g^2 _{SU(N \!+\! 1)}
\left[
{\displaystyle \frac{1}{4}}
\left\{ \!
\left( \!
1 \!-\! {\displaystyle \frac{1}{f^2}} \!
\right) \! N
\!+\!
\left( \!
1 \!+\! 3 {\displaystyle \frac{1}{f^2}} \!
\right) \!
\right\}
\!-\! {\displaystyle \frac{1}{f^2}}
\left\{
(N \!-\! 1) Z^2 \!+\! 1
\right\}
< \!\chi_{\!f}\! >
\right]
\!=\! 0 ,
\EA
\label{VvariZf}
\eeqa
\vspace{-0.3cm}
\beqa
\BA{rl}
&\!\!\!\!
g^2 _{U(1)}
\left[
\xi \!+\! 1
\!+\!
\left( \! 1 \!-\! 2 {\displaystyle \frac{1}{f^2}} \! \right) \!
N
\!+\!
2
{\displaystyle \frac{1}{f^2}}(2 Z^2 \!-\! 1)
< \!\chi_{\!f}\! >
\right] \!
(2 Z^2 \!-\! 1) \\
\\[-10pt]
&\!\!\!\!
-
2 
g^2 _{SU(N \!+\! 1)}
\left[
{\displaystyle \frac{1}{2}}
\left\{ \!
\left( \!
1 \!-\! {\displaystyle \frac{1}{f^2}} \!
\right) \! N
\!+\!
\left( \!
1 \!+\! 3 {\displaystyle \frac{1}{f^2}} \!
\right) \!
\right\} \!
Z^2
\!-\!
{\displaystyle \frac{1}{2}}
\left( \!
1 \!-\! {\displaystyle \frac{1}{f^2}} \!
\right) \! N
\!-\!
{\displaystyle \frac{1}{2}}
\left( \!
1 \!+\! {\displaystyle \frac{1}{f^2}} \!
\right)
\right.\\
\\[-14pt]
&\!\!\!\!
\left.
~~~~~~~~~~~~~~~~~~~~~~~~~
\!-\! {\displaystyle \frac{1}{f^2}}
\left\{
2 (N \!-\! 1) Z^4 \!+\! 4 Z^2 \!-\! (N \!+\! 2)
\right\}
< \!\chi_{\!f}\! >
\right]
\!=\! 0 ,
\EA
\label{Vvarichif}
\eeqa\\[-8pt]
from which, 
we reach to solutions
for $Z^2$ and $< \!\!\chi_{\!f}\!\! >$
obtained in I as\\[-12pt]
\beqa
\!\!\!\!\!\!\!\!\!\!\!\!
\left.
\BA{rl}
&Z^2 
\!=\!
1
\!+\!
{\displaystyle \frac{1 \!-\! f^2}{4 \! < \chi_{\!f} >}}, ~
< \chi_{\!f} >
=\!
{\displaystyle
\frac{1}{2}
\frac{g^2 _{U(1)} \!\!
\left\{ \!
(2 \!-\! f^2) N \!-\! 1 \!
\right\} 
\!-\! 
g^2 _{SU(N \!+\! 1)} \!\!
\left\{ \!
(1 \!-\! f^2) N \!-\! 2 \!
\right\}
\!-\!
g^2 _{U(1)} f^2 \xi } 
{g^2 _{U(1)} \!+\! N g^2 _{SU(N \!+\! 1)}}
} , \\
\\[-12pt]
&\mbox{{\bf tr}}(q^\dag q)
\!=\!
-
{\displaystyle \frac{N}{f^2}}
\left( \!
1 \!-\! {\displaystyle \frac{1}{< \chi_{\!f} >}} \!
\right) , ~
\left( f^2 \! \ge \! 1 \right) .
\EA \!\!\!\!
\right\}
\label{solutionforZ2chif}
\eeqa\\[-10pt]
The third equation in
(\ref{solutionforZ2chif})
is
a vacuum expectation value, i.e.,
$\mbox{{\bf tr}}(q^\dag q)$,
the invariant norm of the complex scalar Goldstone fields.
Putting
(\ref{solutionforZ2chif})
into
(\ref{specialscalerpot2f}),
the minimization of the reduced scalar potential with respect to 
the Fayet-Ilipoulos parameter $\xi$ is realized
as follows:\\[-16pt]
\beqa
\!\!\!\!\!\!\!\!
\left.
\BA{rl}
&\!\!\!\!
V_{\!f{\mbox{\scriptsize redSC}}}
=
{\displaystyle \frac{1}{2}}
{\displaystyle \frac{N}{N \!+\! 1}}
{\displaystyle
\frac{g^2 _{U(1)}g^2 _{SU(N \!+\! 1)}}{g^2 _{U(1)} \!+\! N g^2 _{SU(N \!+\! 1)}}
} \!
\left\{ \!
\xi 
\!-\! {\displaystyle \frac{2 \!-\!\! f^2}{f^2}} \!
N
\!+\!
1
\!+\!
2 {\displaystyle \frac{1}{f^2}}
{\displaystyle \frac{1}{N}} \!
\right\}^{\!2}
\!+\!
V_{\!f{\mbox{\scriptsize redSC}}}^{\mbox{\scriptsize min}} , \\
\\[-6pt]
&\!\!\!
V_{\!f{\mbox{\scriptsize redSC}}}^{\mbox{\scriptsize min}}
\!=\!
{\displaystyle
2 
\frac{g^2 _{SU(N \!+\! 1)}}{N \!+\! 1}
}
{\displaystyle \frac{1}{f^4}} \!
\left[ \!
\left\{ \!
1
\!+\!
{\displaystyle \frac{(1 \!-\!\! f^2)^2}{8}} \!
\right\} \!\! N
\!+\!
{\displaystyle \frac{(1 \!-\!\! f^2)^2}{8}} \!
-\!
{\displaystyle \frac{1}{N}} \!
\right] \! , ~
\xi_{\mbox{\scriptsize min}} 
\!=\!
{\displaystyle \frac{2 \!-\!\! f^2}{f^2}} \!
N
\!-\!
1
\!-\!
2 {\displaystyle \frac{1}{f^2}}
{\displaystyle \frac{1}{N}} . \!
\EA \!
\right\}
\label{minimumscalerpot2f}
\eeqa\\[-10pt]
Putting the
$\xi_{\mbox{\scriptsize min}}$
into 
(\ref{solutionforZ2chif}),
we have the final solutions just the same ones in I as \\[-14pt]
\beqa
\!\!\!\!
Z_{\mbox{\scriptsize min}}^2
\!=\! 
{\displaystyle \frac{1}{2}}
\!+\!
{\displaystyle 
\frac{1}
{2N \! < \!\! \chi_{\!f} \!\! >_{\mbox{\scriptsize min}}}
} , ~~
< \!\! \chi_{\!f} \!\! >_{\mbox{\scriptsize min}}
=\!
{\displaystyle \frac{2 \!-\! N \! (1 \!-\! f^2)}{2N}} , ~~
\mbox{{\bf tr}}(q^\dag q)_{\mbox{\scriptsize min}}
\!=\!
{\displaystyle \frac{N}{f^2}} \!
\left( \!
{\displaystyle
\frac{1}
{< \!\! \chi_{\!f} \!\! >_{\mbox{\scriptsize min}}}
}
\!-\! 1 \!
\right) .
\label{solutionforZ2chifmini}
\eeqa


\newpage


\setcounter{equation}{0}
\renewcommand{\theequation}{\arabic{section}.\arabic{equation}}

\section{Density matrix derived from optimized solutions and vacuum function for bosonized fermions}


~~
Parallel to
(\ref{densitymat}),
let us introduce the following 
$2N \!\times\! 2N$ generalized density matrix:
\beq
W
\!=\!
\left[ 
\BA{cc} 
R & K \\
\\[-8pt]
- \overline{K} & 1_{\!N} \!-\! \overline{R}
\EA 
\right] ,~~
\BA{c}
R
\!=\!
q\overline{\chi } q^\dag ,\\
\\[-8pt]
\!\!\!\!\!
K
\!=\!
\chi q ,
\EA
~W^2 \!=\! W ~(\mbox{idempotency relation}) .
\label{densitymat0}
\eeq
Define a factorized density-matrix 
$< \!\! W \!\! >_{\!f\mbox{\scriptsize min}}$
and use
the optimized $f$-deformed solution
(\ref{solutionforZ2chifmini}):
\beqa
\!\!\!\!
\begin{array}{ll}
< \!\! W \!\! >_{\!f\mbox{\scriptsize min}}
\!\!&\stackrel{\mathrm{def}}{=}\!\!
\left[ \!\!
\BA{cc} 
< \!\! \overline{\chi }_{\!f} \!\! >_{\mbox{\scriptsize min}}
< \!\! q q^\dag \!\! >_{\!f\mbox{\scriptsize min}} \!\cdot 1_{\!N} &\!\!\!\!
< \!\! \chi_{\!f} \!\! >_{\mbox{\scriptsize min}}
< \!\! q \!\! >_{\!f\mbox{\scriptsize min}} \!\cdot 1_{\!N} \\
\\
< \!\! \overline{\chi }_{\!f} \!\! >_{\mbox{\scriptsize min}}
< \!\! \overline{q} \!\! >_{\!f\mbox{\scriptsize min}} \!\cdot 1_{\!N} & \!\!\!\!
1_{\!N} -\! < \!\! \chi_{\!f} \!\! >_{\mbox{\scriptsize min}}
< \!\! \overline{q} q^{\mbox{\scriptsize T}} \!\! >_{\!f\mbox{\scriptsize min}}
\!\cdot 1_{\!N}
\EA \!\!
\right] \\
\\
\!\!&=\!
\left[ \!\!\!
\BA{cc}
< \!\! \chi_{\!f} \!\! >_{\mbox{\scriptsize min}} \!
\left( \!
{\displaystyle
\frac{1}
{< \!\! \chi_{\!f} \!\! >_{\mbox{\scriptsize min}}}
}
- 1 \!
\right)
\!\cdot\! 1_{\!N} & \!\!\!\!
< \!\! \chi_{\!f} \!\! >_{\mbox{\scriptsize min}} \!\!
\sqrt{ \!
{\displaystyle
\frac{1}
{< \!\! \chi_{\!f} \!\! >_{\mbox{\scriptsize min}}}
}
- 1 \!
}
\cdot\! 1_{\!N} \\
\\
< \!\! \chi_{\!f} \!\! >_{\mbox{\scriptsize min}} \!\!
\sqrt{ \!
{\displaystyle
\frac{1}
{< \!\! \chi_{\!f} \!\! >_{\mbox{\scriptsize min}}}
}
- 1 \!
}
\cdot\! 1_{\!N} & \!\!\!
1_{\!N}
-\!
< \!\! \chi_{\!f} \!\! >_{\mbox{\scriptsize min}} \!
\left( \!
{\displaystyle
\frac{1}
{< \!\! \chi_{\!f} \!\! >_{\mbox{\scriptsize min}}}
}
- 1 \!
\right)
\!\cdot\! 1_{\!N}
\EA \!\!\!
\right] \! .
\end{array}
\label{expectdensitymatf0}
\eeqa
In
(\ref{expectdensitymatf0})
the quantities
$< \!\! q q^\dag \!\! >_{\!f\mbox{\scriptsize min}}$
and
$< \!\! q \!\! >_{\!f\mbox{\scriptsize min}}$
are defined
as
$
< \!\! q q^\dag \!\! >_{\!f\mbox{\scriptsize min}}
=\!\!
f^2 \mbox{{\bf tr}}(q^\dag q)_{\mbox{\scriptsize min}}/N
$
(\ref{solutionforZ2chifmini})
and
$
< \!\! q \!\! >_{\!f\mbox{\scriptsize min}}
\!=\!
\sqrt{< \!\! q q^\dag \!\! >_{\!f\mbox{\scriptsize min}}}
$,
respectively.
After calculating square of 
$< \!\! W \!\! >_{\!f\mbox{\scriptsize min}}$,
then we have
\beqa
\!\!\!\!
\begin{array}{ll}
< \!\! W \!\! >_{\!f\mbox{\scriptsize min}}^2
=\!
\left[ \!\!\!
\BA{cc}
< \!\! \chi_{\!f} \!\! >_{\mbox{\scriptsize min}} \!
\left( \!
{\displaystyle 
\frac{1}
{< \!\! \chi_{\!f} \!\! >_{\mbox{\scriptsize min}}}
}
- 1 \!
\right) 
\cdot\! 1_{\!N} & \!\!\!\!
< \!\! \chi_{\!f} \!\! >_{\mbox{\scriptsize min}} \!\!
\sqrt{ \!
{\displaystyle 
\frac{1}
{< \!\! \chi_{\!f} \!\! >_{\mbox{\scriptsize min}}}
}
- 1 \!
} 
\cdot\! 1_{\!N} \\
\\
< \!\! \chi_{\!f} \!\! >_{\mbox{\scriptsize min}} \!\!
\sqrt{ \!
{\displaystyle 
\frac{1}
{< \!\! \chi_{\!f} \!\! >_{\mbox{\scriptsize min}}}
}
- 1 \!
} 
\cdot\! 1_{\!N} & \!\!\!
1_{\!N}
-\!
< \!\! \chi_{\!f} \!\! >_{\mbox{\scriptsize min}} \!
\left( \!
{\displaystyle
\frac{1}
{< \!\! \chi_{\!f} \!\! >_{\mbox{\scriptsize min}}}
}
- 1 \!
\right)
\!\cdot\! 1_{\!N} 
\EA \!\!\!
\right] \!
\!=
< \!\! W \!\! >_{f\mbox{\scriptsize min}} ,
\end{array}
\label{expectdensitymatf02}
\eeqa
which shows that the idempotency relation does hold.
The vacuum function $\Phi_{00}(g)$
in $g \!\in\! SO(2N)$
(\ref{Bogowf})
satisfies\\[-12pt]
\beq
\left(
\mbox{\boldmath $e^\alpha_{~\beta }$} \!+\! \frac{1}{2}\delta_{\alpha \beta }
\right)
\Phi_{00}(g)
\!=\!
\mbox{\boldmath $e_{\alpha \beta }$}\Phi_{00}(g)
\!=\!
0 ,~~
\Phi_{00}(1_{2N})
\!=\!
1 ,
\label{vacuumcondition3}
\eeq
where
$
\mbox{\boldmath $e^\alpha_{~\beta }$},~\mbox{\boldmath $e_{\alpha \beta }$}
~\mbox{and}~
\mbox{\boldmath $e^{\alpha \beta }$}
$,
given in I
\cite{SJCF.08},
are bosonized operators of the $SO(2N)$
fermion Lie operators.

Using the famous formula
$
\det (1_N \!+\! X)
\!=\!
\exp [\mbox{{\bf tr}} \ln (1_N \!+\! X)]
\!=\!
\exp
\left[
\sum_{n=1}^{\infty }
{\displaystyle \frac{(-1)^{n-1}}{n}}\mbox{{\bf tr}}(X)^n
\right]
$,
$\det (1_N \!+\! f^2 q^\dag q)$
is calculated approximately as follows:
\beqa
\!\!\!\!\!\!\!\!
\BA{rl}
&\det (1_N \!+\! f^2 q^\dag q)
\!=\!
\exp
\left[ \!
\sum_{n=1}^{\infty }
{\displaystyle \frac{(-1)^{n-1}}{n}}
\mbox{{\bf tr}} \!
\left\{ \!
\left( \! f^2 q^\dag q \! \right)^n \!
\right\} \!
\right] 
\!\approx\!
\exp
\left[ \!
\sum_{n=1}^{\infty }
{\displaystyle \frac{(-1)^{n-1}}{n}} N^n \!
\left\{ \!
{\displaystyle \frac{1}{N}}
f^2 \mbox{{\bf tr}} \!
\left( \! q^\dag q \! \right) \!
\right\}^n
\right] \\
\\[-8pt]
&\!=\!
\left[ \!
\sum_{n=1}^{\infty }
{\displaystyle \frac{(-1)^{n-1}}{n}} \!
\left\{ \!
N \!
\left( \!
{\displaystyle
\frac{1}
{< \!\! \chi_{\!f} \!\! >}
}
\!-\!
1 \!
\right) \!
\right\}^n
\right]
\!=\!
\exp
\left[
\ln
\left\{ \!
1
\!+\!
N \!
\left( \!
{\displaystyle
\frac{1}
{< \!\! \chi_{\!f} \!\! >}
}
\!-\!
1 \!
\right) \!
\right\}
\right] 
\!=\!
1
\!+\!
N \!
\left( \!
{\displaystyle
\frac{1}
{< \!\! \chi_{\!f} \!\! >}
}
\!-\!
1 \!
\right) .
\EA
\nonumber
\label{SO2Nvacuumfmini}
\eeqa
Then putting the optimized $f$-deformed solution
(\ref{solutionforZ2chifmini})
into the above,
finally we obtain
\beq
\Phi_{00}(g)
\!=\!
\left[
\det (1_N \!+\! f^2 q^\dag q)
\right]^{-\frac{1}{4}} \!
e^{-i\frac{\tau }{2}}
\!=\!
\left[
1
\!+\!
N \!
\left( \!
{\displaystyle
\frac{1}
{< \!\! \chi_{\!f} \!\! >}
}
\!-\!
1 \!
\right) \!
\right]^{-\frac{1}{4}} \!
e^{-i\frac{\tau }{2}} .
\label{SO2Nvacuumfmini2}
\eeq
We should emphasize that a beautiful formula for vacuum function is explicitly derived.


\newpage

\setcounter{equation}{0}
\renewcommand{\theequation}{\arabic{section}.\arabic{equation}}

\section{Anomaly-free $\frac{SO(2N\!+\!2)}{U(N\!+\!1)}$ supersymmetric $\sigma$-model}

~~~
The Lagrangian
${\cal L}_{{\mbox{\scriptsize chiral}}}$
(\ref{susylaglangian})
is invariant under a $U(1)$ symmetry, i.e.,
under multiplication of the superfield
$\phi^{[\alpha]}$
by a universal phase factor $e^{i\widehat{\theta }}$.
According to Nibbelink and van Holten
\cite{NH.98},
the symmetry is expressed in terms of
a holomorphic Killing vectror
${\cal R}^{[\alpha]}_{\widehat{\theta }}({\cal Q})$
by the transformations
\beqa
\left.
\begin{array}{c}
\delta_{\widehat{\theta }}{\cal Q}^{[\alpha]}
\!=\!
\widehat{\theta }{\cal R}^{[\alpha]}_{\widehat{\theta }}({\cal Q})
\!=\!
i\widehat{\theta }q_{([\alpha])}{\cal Q}^{[\alpha]} , \\
\\
\delta_{\widehat{\theta }}\psi^{[\alpha ]}_L
\!=\!
\widehat{\theta }{\cal R}^{[\alpha]}_{\widehat{\theta },~[\beta ]}
({\cal Q})
\psi^{[\beta ]}_L
\!=\!
i\widehat{\theta }q_{([\alpha])}
\psi^{[\alpha ]}_L ,
\end{array}
\right\}
\label{infinitesimaltrans2}
\eeqa
in which
the quantity $q_{([\alpha])}$
means the $U(1)$ charges of the superfields.
There is a larger set of holomorphic Killing vectror
${\cal R}^{[\alpha]}_{\underline{i}} ({\cal Q})$
defining a Lie algebra with structure constants
$f_{\underline{i}\underline{j}}^{~~\underline{k}}$:
\beqa
{\cal R}^{[\beta]}_{\underline{i}} ({\cal Q})
{\cal R}^{[\alpha]}_{{\underline{j}},~[\beta ] }({\cal Q})
\!-\!
{\cal R}^{[\beta]}_{\underline{j}} ({\cal Q})
{\cal R}^{[\alpha]}_{{\underline{i}},~[\beta ] }({\cal Q})
\!=\!
f_{\underline{i}\underline{j}}^{~~\underline{k}}
{\cal R}^{[\alpha]}_{\underline{k}} ({\cal Q}) .
\label{Liealgebrawithstructureconstantf}
\eeqa
Then the Lagrangian
(\ref{susylaglangian})
is invariant under the infinitesimal transformations generated
by the derivation
$\delta
\!=\!
\widehat{\theta }^{\underline{i}}
\delta_{\underline{i}}
$:\\[-16pt]
\beqa
\left.
\begin{array}{c}
\delta{\cal Q}^{[\alpha]}
\!=\!
\widehat{\theta }^{\underline{i}} {\cal R}^{[\alpha]}_{\underline{i}}
({\cal Q}) , ~~
\delta \overline{\cal Q}^{[\underline{\alpha }]}
\!=\!
\widehat{\theta }^{\underline{i}}\overline{{\cal R}}^{[\underline{\alpha }]}
_{\underline{i}}
(\overline{\cal Q}) , \\
\\
\delta\psi^{[\alpha ]}_L
\!=\!
\widehat{\theta }^{\underline{i}}{\cal R}^{[\alpha]}_{{\underline{i}},~[\beta] }
({\cal Q})
\psi^{[\beta ]}_L , ~~
\delta\bar{\psi }^{[\underline{\alpha }]}_L
\!=\!
\widehat{\theta }^{\underline{i}}\overline{{\cal R}}^{[\underline{\alpha }]}
_{{\underline{i}},~[\underline{\beta }]}
(\overline{\cal Q})
\bar{\psi }^{[\underline{\beta }]}_L .
\end{array}
\right\}
\label{infinitesimaltrans3}
\eeqa
Notice that
the K\"{a}hler potential
${\cal K}({\cal Q}^\dag,{\cal Q})$
cannot be determined uniquely
since the metric tensor 
${\cal G}_{pq}{}_{\underline{r}\underline{s}}$
is invariant under transformations of the K\"{a}hler potential,
\beq
{\cal K}({\cal Q}^\dag,{\cal Q})
\!\rightarrow\!
{\cal K}^\prime ({\cal Q}^\dag,{\cal Q})
\!=\!
{\cal K}({\cal Q}^\dag,{\cal Q})
\!+\! {\cal F}({\cal Q})
\!+\! \overline{\cal F}(\overline{\cal Q}) .
\label{transKpot}
\eeq
Under the holomorphic transformations
(\ref{infinitesimaltrans3})
the K\"{a}hler potential itself transforms as
\beq
\delta_{\underline{i}} {\cal K}({\cal Q}^\dag,{\cal Q})
\!=\!
{\cal F}_{\underline{i}} ({\cal Q})
\!+\!
\overline{\cal F}_{\underline{i}} (\overline{\cal Q}) ,
\label{delKpot}
\eeq
where
${\cal F}_{\underline{i}} ({\cal Q})$
and
$\overline{\cal F}_{\underline{i}} (\overline{\cal Q})$
are analytic functions of
${\cal Q}$ and $\overline{\cal Q}$, respectively
and related to a set of real scalar potentials
${\cal M}_{\underline{i}} ({\cal Q},\overline{\cal Q})$
satisfying
$
\delta_{\underline{i}} {\cal M}_{\underline{i}} ({\cal Q},\overline{\cal Q})
\!=\!
f_{{\underline{i}}{\underline{j}}}^{~~{\underline{k}}}
{\cal M}_{\underline{k}} ({\cal Q},\overline{\cal Q})
$
as
\beqa
\left.
\begin{array}{c}
{\cal F}_{\underline{i}} ({\cal Q})
\!=\!
{\cal K}({\cal Q}^\dag,{\cal Q})_{,[\alpha]}
{\cal R}^{[\alpha]}_{\underline{i}} ({\cal Q})
\!+\!
i {\cal M}_{\underline{i}} ({\cal Q},\overline{\cal Q}) ,\\
\\
\overline{{\cal F}}_{\underline{i}} (\overline{{\cal Q}})
\!=\!
{\cal K}({\cal Q}^\dag,{\cal Q})_{,[\underline{\alpha]}}
\overline{{\cal R}}^{\underline{[\alpha]}}_{\underline{i}}
(\overline{{\cal Q}})
\!+\!
i {\cal M}_{\underline{i}} ({\cal Q},\overline{\cal Q}) .
\end{array}
\right\}
\label{holomorFi}
\eeqa
The pure SUSY $\sigma$-models on cosets
including Grassmannian models
on $\frac{SU(N + M)}{[SU(N) \times SU(M) \times U(1)]}$
and
models on manifolds $\frac{SO(2N)}{U(N)}$ are recognized to be anomalous,
because they incorporate chiral fermions in non-trivial representations
of the holonomy group
\cite{NNH.01}.
The presence of the chiral anomalies in the internal symmetry
(\ref{infinitesimaltrans3})
restricts the usefulness of these models for phenomenological applications.
These anomalies must be removed to allow for a consistent gauging of
the symmetries, e.g., the chiral $U(1)$ symmetry
(\ref{infinitesimaltrans2}).
It is realized by coupling additional chiral fermions
to the $\sigma$-model
preserving the holomorphic Killing vectors.

Generally speaking,
if one constructs some quantum field theories
based on pure coset models,
one has with serious problems of anomalies
in a holonomy group
which particularly occur in pure SUSY coset models
due to the additional chiral fermions.
This is also the cases for our orthogonal cosets,
coset $\frac{SO(2N)}{U(N)}$ and
its extended coset $\frac{SO(2N\!+\!2)}{U(N\!+\!1)}$,
though each spinor rep of $SO(2N)$ group and
its extended $SO(2N \!\!+\!\! 2)$ group
is anomaly free.
To construct a consistent SUSY coset model,
we have to embed a coset coordinate
in an anomaly-free spinor rep of $SO(2N \!\!+\!\! 2)$ group
and give a corresponding K\"{a}hler potential and
then a Killing potential for the anomaly-free
$\frac{SO(2N\!+\!2)}{U(N\!+\!1)}$ model
based on a positive chiral spinor rep.
To achieve such an object on the case of $SO(2N)$ group/algebra,
van Holten et al. have proposed
a method of constructing
the K\"{a}hler potential and then the Killing potential
\cite{NNH.01}.
This idea is very suggestive and useful for our present aim of constructing
the corresponding K\"{a}hler potential and then the Killing potential
for the case of $SO(2N \!\!+\!\! 2)$ group/algebra.

According to Fukutome
\cite{Fuk.81}
in the $SO(2N \!\!+\!\! 2)$ Lie algebra
the total fermion space is irreducible to the $SO(2N \!\!+\!\! 2)$ algebra and belongs to the irreducible spinor rep of the $SO(2N \!\!+\!\! 2)$ group.
It is well known that
the dimension of irreducible spinor rep of
the $SO(2N \!\!+\!\! 2)$ group is $2^N$,
so that the $SO(2N \!\!+\!\! 2)$ algebra can be accomodated
in the fermion space.
The Clifford algebra $C_{2N\!+\!2}$ is defined on a space
with $2^{N\!+\!1}$ dimensions,
so that it cannot be constructed on the fermion space though
the $SO(2N \!\!+\!\! 2)$ algebra with irreducible spinor rep of
$2^N$ dimensions can be accomodated in the fermion space.
The operators
$E^i_{~j},~E^{ij}$ and $E_{ij}$
satisfy
the commutation relations of the $SO(2N \!\!+\!\! 2)$ Lie algebra
where
the indices $i,j,\dots$ run over
$N \!+\! 1$ values $0,1,\dots,N$.
The operator
$
E^i_{~i} \!+\! \frac{N}{2}
\!=\!
n \!-\! \frac{1}{2} (-1)^n
$
is the only operator commuting with all other operators 
in the $U(N \!\!+\!\! 1)$ algebra.
The $U(1)$-factor generator $Y$ in $U(N)$ is defined as
$
Y
\!=\!
2 E^i_{~i}
\!=\!
2n \!-\! N \!-\! (-1)^n
$
and the remaining $SU(N \!+\! 1)$ generators
$T^i_{~i}$
are defined as the traceless part of
$E^i_{~i}$,
$
T^i_{~i}
\!=\!
E^i_{~i}
\!+\!
\frac{1}{2(N \!+\! 1)} Y
\delta ^i_{~j}
$.
Extending the van Holten et al.'s formula for $Y^k$-anomaly
from the $SO(2N)$ case
\cite{NNH.01}
to the $SO(\!2N \!+\! 2\!)$ case,
we also define a new $A_{\pm }(\! Y^k;N\!+\!1 \!)$-anomaly as
$
A_{\pm }(\! Y^k;N\!+\!1 \!)
\!=\!\!
\sum_{m\!=\!0}^{N\!+\!1} \!
\binom{N\!+\!1}{m} 
\frac{1 \!\pm\! (\! -1 \!)^m}{2} \!
\left\{ \!
N \!-\! 2m \!-\! (-1)^m \!
\right\}^{k}
$.
This formula makes an important role to caculate the $U(1)$-anomaly
in the spinor rep of the
$SO(\!2N \!+\! 2\!)$.
By using the $SO(2N \!\!+\!\! 2)$ Lie operators $E^{ij}$,
the expression
(\ref{SO2Nplus1wf})
for the $SO(2N \!\!+\!\! 1)$ wave function $\ket G$ is converted to a form quite similar to the $SO(2N)$ wave function $\ket g$ as
\beq
\ket G
\!=\!
\bra 0 U(G) \ket 0
\exp \! \left( \! 1 / 2 \cdot {\cal Q}_{ij}E^{ij} \! \right) \! \ket 0 ,
\label{SO2N+1wf}
\eeq
which leads to
$U(G) \ket 0 \!=\! U({\cal G}) \ket 0$
and
we have used the nilpotency relation $(E^{\alpha 0})^2 \!=\! 0$.
The construction of the K\"{a}hler potential and then the Killing potential for the $SO(\!2N\!+\!2\!)$ group/algebra
is made parallel to the construction of those
for the $SO(2N)$ group/algebra.

First according to
\cite{NNH.01}
we define a matrix
$\Xi ({\cal Q})$ and require a transformation rule as follows:\\[-2pt]
\beq
\Xi ({\cal Q})
\!\stackrel{\mathrm{def}}{=}\!
\left[ \!\!
\BA{cc}
1_{N \!+\! 1} & 0 \\
\\[-6pt]
{\cal Q} & 1_{N \!+\! 1} \\
\EA \!\!
\right] ,
\label{matXi}
\eeq
\vspace{-0.3cm}
\beqa
\!\!\!\!
\Xi ({\cal Q})
\!\longrightarrow\!
\Xi ({}^{{\cal G}} \! {\cal Q})
\!=\!
{\cal G} \Xi ({\cal Q}) \widehat{H}^{-1}({\cal Q}; {\cal G}),
~\mbox{with}~
\widehat{H} ({\cal Q}; {\cal G})
\!=\!
\left[ \!\!
\BA{cc}
\left(\! \widehat{H}_+ ({\cal Q}; {\cal G} \! \right)^{-1} &
\widehat{H}_0 ({\cal Q}; {\cal G}) \\
\\[-6pt]
0 & \widehat{H}_- ({\cal Q}; {\cal G})
\EA \!\!
\right] \! ,
\label{transmatXi}
\eeqa\\[-12pt]
where
\beqa
\!\!\!\!
{}^{{\cal G}} \! {\cal Q}
\!=\!
({\cal B} \!+\! \bar{\cal A} {\cal Q})
({\cal A} \!+\! \bar{\cal B} {\cal Q})^{-1}
\!=\!
({\cal A}^{\mbox{\scriptsize T}} \!-\! {\cal Q}{\cal B}^\dag )^{-1}
({\cal B}^{\mbox{\scriptsize T}} \!-\! {\cal Q}{\cal A}^\dag ) ,~
\mbox{(due to ${\cal G}^{-1} \!=\! {\cal G}^{\dag }$)} .
\label{MtransmatQ}
\eeqa
The
$
{}^{{\cal G}} \! {\cal Q}
$
is a nonlinear M\"{o}bius transformation
and 
$
{}^{{\cal G}^\prime } \! ({}^{{\cal G}} \! {\cal Q})
\!=\!
{}^{{\cal G}^\prime }{}^{{\cal G}} \! {\cal Q}
$
under the composition of two transformations
${\cal G}^\prime$ and ${\cal G}$.
Under the action of the $SO(2N\!+\!2)$ matrix ${\cal G}$
(\ref{calG}) on $\Xi ({\cal Q})$ from the left and 
that of the matrix $\widehat{H} ({\cal Q}; {\cal G})^{-1}$ from the right,
the
$\widehat{H} ({\cal Q}; {\cal G})$
takes the form
\beq
\widehat{H} ({\cal Q}; {\cal G})
\!=\!\!
\left[ \!\!\!
\BA{cc}
\left(\! \widehat{H}_+ ({\cal Q}; {\cal G} \! \right)^{\!\!-1} &\!\!\!
\widehat{H}_0 ({\cal Q}; {\cal G}) \\
\\[-16pt]
0 &\!\!\! \widehat{H}_- ({\cal Q}; {\cal G}) \\
\EA \!\!\!
\right]
\!\!=\!\!
\left[ \!\!\!
\BA{cc}
{\cal A} \!+\! \bar{\cal B} {\cal Q} &\!\!\! \bar{\cal B} \\
\\[-4pt]
0 &\!\!\! ({\cal A}^{\mbox{\scriptsize T}}
\!-\!
{\cal Q}{\cal B}^\dag )^{\!-1}
\EA \!\!\!
\right] \! ,~
\widehat{H}_+ ({\cal Q}; {\cal G})
\!=\!
\widehat{H}_- ^{\mbox{\scriptsize T}} ({\cal Q}; {\cal G}) .
\label{hatHform}
\eeq
Then we have
$
\det \widehat{H}_+ ({\cal Q}; {\cal G})
\!=\!
\det \widehat{H}_- ({\cal Q}; {\cal G})
$.
Multiplying ${\cal G}^\prime$ by ${\cal G}$,
we have the relations\\[-10pt]
\beq
\widehat{H}_- ({\cal Q}; {\cal G}^\prime {\cal G})
\!=\!
\widehat{H}_- ({}^{{\cal G}} \! {\cal Q}; {\cal G}^\prime)
\widehat{H}_- ({\cal Q}; {\cal G}) , ~~
\widehat{H}_+ ({\cal Q}; {\cal G}^\prime {\cal G})
\!=\!
\widehat{H}_+ ({\cal Q}; {\cal G})
\widehat{H}_+ ({}^{{\cal G}} \! {\cal Q}; {\cal G}^\prime) .
\label{decompoHminus}
\eeq
Here we redefine the K\"{a}hler potential as
$
{\cal K}({\cal Q},\overline{{\cal Q}})
\!\!=\!\!
\ln \det \!
\left(
1_{N \!+\! 1}
\!+\!
{\cal Q} \overline{{\cal Q}}
\right)
$.
Under the nonlinear transformation
(\ref{MtransmatQ}),
the K\"{a}hler potential
transforms as\\[-8pt]
\beq
{\cal K} ({}^{{\cal G}} \! {\cal Q},
{}^{{\cal G}} \overline{{\cal Q}})
\!=\!
{\cal K}({\cal Q} ,\overline{{\cal Q}})
\!+\!
{\cal F}({\cal Q} ;{\cal G})
\!+\!
\overline{{\cal F}}
(\overline{{\cal Q}};{\cal G}) ,
\label{MtransKpot}
\eeq
which holds for any coordinate ${\cal Q}$
and
any frame
${\cal G}$.
Then we have an approximate relation
\beq
{\cal F} ({\cal Q} ;{\cal G} )
\!=\!
\ln \det \widehat{H}_- ({\cal Q} ; {\cal G})
\!=\!
-
\ln \det
\left[
{\cal A}^{\mbox{\scriptsize T}}
\!\!-\!\!
{\cal Q} {\cal B}^{\dag }
\right]
\!=\!
-
\mbox{tr} \ln 
\left[
{\cal A}^{\mbox{\scriptsize T}}
\!\!-\!\!
{\cal Q} {\cal B}^{\dag }
\right]
\!\approx\!
\mbox{tr} [{\cal R}_T ({\cal Q}; \delta {\cal G})] ,
\label{holomorF2}
\eeq
where
we have expanded
$
{\cal A}^{\mbox{\scriptsize T}}
~\mbox{and}~
{\cal B}^{\dag }
$
to the first order in the infinitesimal paramerters
$\delta {\cal A}^{\mbox{\scriptsize T}}$
and
$\delta {\cal B}^\dag$
and have used the second equation of
(\ref{RRTChi}).

Next we add matter superfields to extend the model on which
an isometry group is realized with a representation chosen
to cancel the anomalies.
A matter representation of the isometry group
is constructed by complex bundles
defined on a K\"{a}hler manifold by sets of
complex fields with the transformation
(\ref{infinitesimaltrans3})
for chiral fermions under isometries.
If one requires anomaly cancellations with the matter superfields,
one may change an assignment of $U(1)$ charges
by introducing a complex line bundle ${\cal S}$.
This bundle can be defined as a complex scalar matter field
coupled to the SUSY $\sigma$-model,
with the infinitesimal transformation law\\[-10pt]
\beq
\delta_{\underline{i}} {\cal S}^\lambda
\!=\!
\lambda {\cal F}_{\underline{i}} ({\cal Q}) {\cal S} .
\label{infinitesimaltransS}
\eeq
For a tensor representation of the isometry group
$
{\cal T}^{\alpha_1 \cdots \alpha_p}
\!\equiv\!
{\cal S}^\lambda T^{\alpha_1 \cdots \alpha_p}
$,
the new field
${\cal T}$
obeys the transformation rule\\[-20pt]
\beqa
\begin{array}{c}
\delta_{\underline{i}} {\cal T}^{\alpha_1 \cdots \alpha_p}
\!=\!
\sum_{k \!=\! 1} ^p
{\cal R}^{\alpha_k}_{{\underline{i}},~\beta }({\cal Q})
{\cal T}^{\alpha_1 \cdots \beta \dots \alpha_p}
\!+\!
\lambda {\cal F}_{\underline{i}} ({\cal Q})
{\cal T}^{\alpha_1 \cdots \alpha_p} .
\end{array}
\label{infinitesimaltransT}
\eeqa\\[-14pt]
A section of a minimal line bundle over
$\frac{SO(2N\!+\!2)}{U(N\!+\!1)}$
is given by\\[-8pt]
\beq
{}^{{\cal G}} \! {\cal S}
\!=\!
\left[
\det \widehat{H}_+ ({\cal Q}; {\cal G})
\right]^{\frac{1}{2}} \! {\cal S}
\!=\!
\left[
\det \widehat{H}_- ({\cal Q}; {\cal G})
\right]^{\frac{1}{2}} \! {\cal S} .
\label{sectionS}
\eeq\\[-12pt]
Suppose that
${\cal T}_{(p;q)}^{i_1 \!\cdots\! i_p}$
is an irreducible completely antisymmetric $SU( \! N \!+\! 1 \!)$-tensor
representation with $p$ indices and arbitrary rescaling charge $q$.
We abbreviate it simply as ${\cal T}_{(p;q)}$.
By taking the completely antisymmetric tensor product of a set of
$SU(N \!+\! 1)$ vectors
$\{ {\cal T}_1 ^{i_1}, \!\cdots\!, {\cal T}_p ^{i_p} \}$
we obtain an $SU( \! N \!+\! 1 \!)$ tensor of rank $p$
with rescaling charge $q$\\[-10pt]
\beq
{\cal T}_{(p;q)} ^{i_1 \cdots i_p}
\!\equiv\!
\frac{1}{p!}
{\cal S}^q T_1 ^{[ i_1} \ast \cdots \ast T_p ^{i_p ]} ,
\label{SUNplus1tensor}
\eeq
where
$[ \cdots ]$
denotes the completely anti-symmetrization of the indices
inside the brackets.
Thus we obtain a transformation of tensor
${\cal T}_{(p;q)} ^{i_1 \cdots i_p}$
as\\[-8pt]
\beq
{}^{{\cal G}}{\cal T}_{(p;q)} ^{i_1 \cdots i_p}
\!=\!
\left[
\det \widehat{H}_- ({\cal Q}; {\cal G})
\right]^{\!\frac{q}{2}} \!
\left[
\widehat{H}_- ({\cal Q}; {\cal G})
\right]_{~~j_1} ^{i_1} \!
\cdots \!
\left[
\widehat{H}_- ({\cal Q}; {\cal G})
\right]_{~~j_p} ^{i_p} \!
{\cal T}_{(p;q)} ^{j_1 \cdots j_p} .
\label{transSUNplus1tensor}
\eeq
The invariant K\"{a}hler potential for a tensor
is given by\\[-16pt]
\beqa
{\cal K}_{(p;q)}
\!=\!
\overline{{\cal T}}_{(p;q) j_1 \cdots j_p}
{\cal G}_{(p;q) i_1 \cdots i_p} ^{j_1 \cdots j_p}
{\cal T}_{(p;q)} ^{i_1 \cdots i_p} ,~
{\cal G}_{(p;q) i_1 \cdots i_p} ^{j_1 \cdots j_p}
\!\equiv\!
\frac{1}{p!}
[\det {\cal X}]^{\frac{q}{2}}
{\cal X}_{~~i_1} ^{j_1}
\cdots
{\cal X}_{~~i_p} ^{j_p} .
\label{KpotSUNplus1tensor}
\eeqa\\[-16pt]
An $SU(N \!+\! 1)$ dual tensor
$
{\cal T}_{(\overline{N \!+\! 1 \!-\! p};q)
i_{p \!+\! 1} \cdots i_{N \!+\! 1}}
$
with $( N \!+\! 1 \!-\! p )$ indices and
rescaling charge $q$ is\\[-12pt]
\beqa
{\cal T}_{(\overline{N \!+\! 1 \!-\! p};q)
i_{p \!+\! 1} \cdots i_{N \!+\! 1}}
\!\equiv\!
\frac{1}{p!}
{\cal T}_{(p;q)} ^{i_p \cdots i_1}
\epsilon_{i_1 \cdots i_{N \!+\! 1}} ,~
(\epsilon_{i_1 \cdots i_{N \!+\! 1}}: SU(N \!+\! 1)
\mbox{Levi-Civita tensor} )
\label{dualSUNplus1tensor}
\eeqa\\[-16pt]
which transforms under the nonlinear M\"{o}bius transformation
(\ref{MtransmatQ})
as\\[-16pt]
\beqa
{}^{{\cal G}}{\cal T}_{(\overline{p};q) i_1 \cdots i_p}
\!=\!
{\cal T}_{(\overline{p};q) j_1 \cdots j_p}
\left[
\widehat{H}_- ^{-1} ({\cal Q}; {\cal G})
\right]_{~~i_1} ^{j_1}
\cdots
\left[
\widehat{H}_- ^{-1} ({\cal Q}; {\cal G})
\right]_{~~i_p} ^{j_p}
\left[
\det \widehat{H}_- ({\cal Q}; {\cal G})
\right]^{1 \!+\! \frac{q}{2}} .
\label{transdualSUNplus1tensor}
\eeqa\\[-12pt]
The invariant K\"{a}hler potential for a dual tensor
is given by\\[-16pt]
\beqa
\begin{array}{l}
{\cal K}_{(\overline{p};q)}
\!=\!
{\cal T}_{(\overline{p};q) i_1 \cdots i_p}
{\cal G}_{(\overline{p};q) j_1 \cdots j_p} ^{i_1 \cdots i_p}
\overline{{\cal T}}_{(\overline{p};q)} ^{j_1 \cdots j_p} ,~
{\cal G}_{(\overline{p};q) j_1 \cdots j_p} ^{i_1 \cdots i_p}
\!\equiv\!
{\displaystyle \frac{1}{p!}}
[\det {\cal X}]^{1 \!+\! \frac{q}{2}}
\left[ {\cal X}^{-1} \right]_{~~j_1} ^{i_1}
\cdots
\left[ {\cal X}^{-1} \right]_{~~j_p} ^{i_p}.
\end{array}
\label{KpotSUNplus1dualtensor}
\eeqa\\[-26pt]

The contributions of the invariant K\"{a}hler potentials
${\cal K}_{(p;q)}$
and
${\cal K}_{(\overline{p};q)}$
to the Killing potentials,
${\cal M}_{(p;q)}(\! {\cal Q},\overline{\cal Q}; \delta {\cal G} \!)$
and
${\cal M}_{(\overline{p};q)}(\! {\cal Q},\overline{\cal Q};
\delta {\cal G} \!)$
for a tensor
${\cal T}_{(\overline{p};q)}$
and a dual tensor
$\overline{{\cal T}}_{(\overline{p};q)}$
of rank $p$ with a rescaling charge $q$,
are obtained from
(\ref{holomorFi})
to satisfy
${\cal F}_i (\! {\cal Q} \!) \!=\! 0$
and
$\overline{{\cal F}}_i (\! \overline{{\cal Q}} \!) \!=\! 0$
as\\[-14pt]
\beqa
\begin{array}{c}
-i {\cal M}_{(\binom{p}{\overline{p}};q)} 
(\! {\cal Q},\overline{\cal Q}; \delta {\cal G} )
=
{\cal K}_{(\binom{p}{\overline{p}};q),~[\alpha]}
(\! {\cal Q},\overline{\cal Q} )
{\cal R}^{[\alpha]} ( {\cal Q}) ,
\end{array}
\label{KllingpotM}
\eeqa\\[-16pt]
where
the infinitesimal transformations
generated by the derivation
$
\delta_{\underline{i}} \phi ^{[\alpha]}
\!=\!
{\cal R}_{\underline{i}} ^{[\alpha]} ( {\cal Q} )
$
denote the Killing vectors
$
\delta {\cal Q}
\!=\!
{\cal R}({\cal Q}; \delta {\cal G}) 
$
and the fields ${\cal T}$ obey the transformation rules\\[-12pt]
\beqa
\begin{array}{l}
\delta {\cal T}_{(p;q)} ^{i_1 \cdots i_p}
\!=\!
\sum_{r \!=\! 1} ^p
\left[
{\cal R}_T ({\cal Q}; \delta {\cal G})
\right]_{~~j} ^{i_r}
{\cal T}_{(p;q)} ^{i_1 \cdots j \cdots i_p}
+
{\displaystyle \frac{q}{2}}
\mbox{tr}
\left[
{\cal R}_T ({\cal Q}; \delta {\cal G})
\right]
{\cal T}_{(p;q)} ^{i_1 \cdots i_p} ,
\end{array}
\label{transKillingtesor}
\eeqa
\vspace{-0.6cm}
\beqa
\!\!\!\!
\begin{array}{l}
\delta {\cal T}_{(\overline{p};q) i_1 \cdots i_p}
\!=\!
\sum_{r \!=\! 1} ^p
{\cal T}_{(\overline{p};q) i_1 \cdots j \cdots i_p}
\left[
- {\cal R}_T ({\cal Q}; \delta {\cal G})
\right]_{~~j_r} ^{j}
+
\left( 1 \!+\! {\displaystyle \frac{q}{2}} \right)
\mbox{tr}
\left[
{\cal R}_T ({\cal Q}; \delta {\cal G})
\right]
{\cal T}_{(\overline{p};q) i_1 \cdots i_p} ,
\end{array}
\label{transKillingbartesor}
\eeqa\\[-12pt]
where we have used
(\ref{holomorF2})
with 
$\lambda \!\!=\!\! q / 2$
in
(\ref{transKillingtesor})
and
(\ref{infinitesimaltransT})
with 
$\lambda \!\!=\!\! 1 \!+\! q / 2$
in
(\ref{transKillingbartesor}),
respectively.
According to van Holten et al.
\cite{NNH.01},
the transformation rules
(\ref{transKillingtesor})
and
(\ref{transKillingbartesor})
also can be
derived by expanding the finite transformations
(\ref{transSUNplus1tensor})
and
(\ref{transdualSUNplus1tensor})
to the first order in the infinitesimal parameters
$\delta {\cal A}^{\mbox{\scriptsize T}}$
and
$\delta {\cal B}^\dag$.
As an example we demonstrate the following variation:\\[-16pt]
\beqa
\!\!\!\!
\delta \!
\left[ \!
\widehat{H}_- ^{\!-1} ({\cal Q}; {\cal G}) \!
\right]_{~j_r} ^{i_r}
\!\!=\!
\left[ \!
\left( \!
{\cal A}^{\mbox{\scriptsize T}}
\!\!-\!\!
{\cal Q} {\cal B}^{\dag }
\right)^{\!-1} \!
\right]_{~j_r} ^{j} \!\!
\left( \!
- \delta {\cal A}^{\mbox{\scriptsize T}}
\!\!+\!\!
{\cal Q} \delta {\cal B}^{\dag }
\right)_{~i} ^{j} \!
\left[ \!
\left( \!
{\cal A}^{\mbox{\scriptsize T}}
\!\!-\!\!
{\cal Q} {\cal B}^{\dag }
\right)^{\!-1} \!
\right]_{~i} ^{i_r}
\!\!\approx\!
\left[
{\cal R}_T ({\cal Q}; \delta {\cal G})
\right]_{~i_r} ^{j_r} \! .
\eeqa
Equation
(\ref{KllingpotM})
is an extended form of$\!$
(\ref{holomorFi})
$\!\!(\!{\cal F}_i \!\!=\!\! \overline{{\cal F}}_i \!\!=\!\! 0 )$
to tensors
and its final expression reads\\[-14pt]
\beqa
\!\!\!\!
\begin{array}{l}
-i {\cal M}_{(p;q)} (\! {\cal Q},\overline{\cal Q}; \delta {\cal G} )
\!=\!
{\displaystyle \frac{1}{p!}}
\overline{{\cal T}}_{(p;q) j_1 \cdots j_p} \\
\\[-8pt]
~~
\times
[{\det {\cal X}}]^{\frac{q}{2}}
{\cal X}_{~~k_1} ^{j_1}
\cdots
{\cal X}_{~~k_p} ^{j_p} \\
\\[-8pt]
~~
\times 
\left\{
\sum_{r \!=\! 1} ^p
\delta_{~~i_1} ^{k_1}
\cdots
\left[ 
\Delta(\! {\cal Q},\overline{\cal Q};
\delta {\cal G} ) 
\right]
_{~~i_r} ^{k_r}
\cdots
\delta_{~~i_p} ^{k_p}
\!+\!
{\displaystyle \frac{q}{2}}
\mbox{tr}
\left[ 
\Delta ( {\cal Q},\overline{\cal Q}; \delta {\cal G} )
\right]
\delta_{~~i_1} ^{k_1}
\cdots
\delta_{~~i_p} ^{k_p} 
\right\}
{\cal T}_{(p;q)} ^{i_1 \cdots i_p} ,
\end{array}
\label{KllingpotM2}
\eeqa
\vspace{-0.3cm}
\beqa
\!\!\!\!\!\!\!\!
\begin{array}{l}
-i {\cal M}_{(\overline{p};q)} (\! {\cal Q},\overline{\cal Q}; \delta {\cal G} )
\!=\!
{\displaystyle \frac{1}{p!}}
{\cal T}_{(\overline{p};q) j_1 \cdots j_p} \\
\\[-8pt]
~~
\times
\left\{ 
\sum_{r \!=\! 1} ^p
\delta_{~~k_1} ^{j_1}
\!\cdots\!
[\! -\Delta (\! {\cal Q},\overline{\cal Q};
\delta {\cal G} )\! ]_{~~k_r} ^{j_r}
\!\cdots\!
\delta_{~~k_p} ^{j_p}
\!+\!
\left(
{\displaystyle \! 1 \!+\! \frac{q}{2}}
\right)
\mbox{tr}
\left[
\Delta (\! {\cal Q},\overline{\cal Q}; \delta {\cal G} )
\right]
\delta_{~~k_1} ^{j_1}
\!\cdots\!
\delta_{~~k_p} ^{j_p}
\right\} \\
\\[-8pt]
~~
\times 
[{\det {\cal X}}]^{1 \!+\! \frac{q}{2}}
[{\cal X}^{-1}]_{~~i_1} ^{k_1}
\cdots
[{\cal X}^{-1}]_{~~i_p} ^{k_p}
\overline{{\cal T}}_{(\overline{p};q)} ^{i_1 \cdots i_p} .
\end{array}
\label{KllingpotM3}
\eeqa
The derivation of equations
(\ref{KllingpotM2}) and (\ref{KllingpotM3})
is made in the following way:
From
(\ref{KllingpotM})
and
(\ref{KpotSUNplus1tensor})
the Killing potential for tensors
is given as
\beqa
\!\!\!\!
\begin{array}{rl}
- i {\cal M}_{(p;q)}
\!\!\!&\!=\!
{\cal K}_{(p;q),~[i]} {\cal R}^{[i]}
\!=\!
\overline{{\cal T}}_{(p;q) j_1 \cdots j_p,~[i]} {\cal R}^{[i]}
{\cal G}_{(p;q) i_1 \cdots i_p} ^{j_1 \cdots j_p}
{\cal T}_{(p;q)} ^{i_1 \cdots i_p} \\
\\[-4pt]
&\!+\!
\overline{{\cal T}}_{(p;q) j_1 \cdots j_p}
{\cal G}_{(p;q) i_1 \cdots i_p,~[i]} ^{j_1 \cdots j_p}
{\cal R}^{[i]}
{\cal T}_{(p;q)} ^{i_1 \cdots i_p}
\!+\!
\overline{{\cal T}}_{(p;q) j_1 \cdots j_p}
{\cal G}_{(p;q) i_1 \cdots i_p} ^{j_1 \cdots j_p}
{\cal T}_{(p;q),~[i]} ^{i_1 \cdots i_p}
{\cal R}^{[i]} ,~([i]\!=\!(i\hat{i})) ,
\end{array}
\label{variKpotSUNplus1tensor}
\eeqa
in which
${\cal Q}^{[i]}$
means the $(i\hat{i})$ element of the matrix
${\cal Q}~(\hat{i}\mbox{: another component different from}~i)$, i.e.,
${\cal Q}_{i\hat{i}}$
and
${\cal R}$ are given by the Killing vector, i.e.,
$
{\cal R}
\!=\!
\delta {\cal Q}
$
(\ref{infinitesimaltrans}).
The variation of
$\delta {\cal G}_{(p;q) i_1 \cdots i_p} ^{j_1 \cdots j_p}$
is calculated as
\beqa
\begin{array}{rl}
\delta {\cal G}_{(p;q) i_1 \cdots i_p} ^{j_1 \cdots j_p}
\!\!\!&=-
{\displaystyle \frac{q}{2} \frac{1}{p!}}
[\det {\cal X}]^{\frac{q}{2}}
\mbox{tr} \!
\left\{ \!
{\cal X} \!
\left(
\delta {\cal Q} \overline{{\cal Q}}
\!+\!
{\cal Q} \delta \overline{{\cal Q}}
\right) \!
\right\}
{\cal X}_{~~i_1} ^{j_1}
\cdots
{\cal X}_{~~i_p} ^{j_p} \\
\\[-8pt]
&-
{\displaystyle \frac{1}{p!}}
[\det {\cal X}]^{\frac{q}{2}} \!
\sum_{r\!=\!1} ^p
{\cal X}_{~~i_1} ^{j_1}
\cdots
\left\{ \!
{\cal X} \!
\left(
\delta {\cal Q} \overline{{\cal Q}}
\!+\!
{\cal Q} \delta \overline{{\cal Q}}
\right) \!
{\cal X} \!
\right\}_{~~i_r} ^{j_r}
\cdots
{\cal X}_{~~i_p} ^{j_p} ,
\end{array}
\label{delGSUNplus1tensor}
\eeqa
together with
$
\delta \det \! {\cal X}
\!\!=\!
- \det {\cal X}
\!\cdot\!
\mbox{tr} \!
\left\{ \!
{\cal X} \!
\left( \!
\delta {\cal Q} \overline{{\cal Q}}
\!+\!
{\cal Q} \delta \overline{{\cal Q}}
\right) \!
\right\}
$
and
$
\delta {\cal X}_{~i}^j
\!=\!
-
\left\{ \!
{\cal X} \!
\left( \!
\delta {\cal Q} \overline{{\cal Q}}
\!+\!
{\cal Q} \delta \overline{{\cal Q}} \!
\right) \!
{\cal X} \!
\right\}_{~i}^j
$.
Taking only the
$\delta {\cal Q}$
term
in
(\ref{delGSUNplus1tensor}),
the following type of contraction is easily carried out:
\beqa
\!\!\!\!
\begin{array}{rl}
{\cal G}_{(p;q) i_1 \cdots i_p,~\hat{i}i} ^{j_1 \cdots j_p}
\delta {\cal Q}_{i\hat{i}}
\!\!\!&=\!
- {\displaystyle \frac{q}{2}\frac{1}{p!}}
[\det {\cal X}]^{\frac{q}{2}}
\mbox{tr} \left( {\cal R}_T \!-\! \Delta \right)
{\cal X}_{~~i_1} ^{j_1}
\cdots
{\cal X}_{~~i_p} ^{j_p} \\
\\[-8pt]
&-
{\displaystyle \frac{1}{p!}}
[\det {\cal X}]^{\frac{q}{2}} \!
\sum_{r\!=\!1} ^p
{\cal X}_{~~i_1} ^{j_1}
\cdots
\left\{ \!
{\cal X}_{~~i} ^{j_r}
\left( {\cal R}_T \!-\! \Delta \right)_{~i_r} ^{i} \!
\right\}
\cdots
{\cal X}_{~~i_p} ^{j_p} ,
\end{array}
\label{sumdelGSUNplus1tensor}
\eeqa
where we have used the relation
$
\delta {\cal Q} \overline{{\cal Q}} {\cal X}
\!=\!
{\cal R}_T - \Delta
$
(\ref{RRTChi}) .
$
\overline{{\cal T}}_{(p;q) j_1 \cdots j_p,[i]} {\cal R}^{[i]}
\!=\!
0
$
is evident
and
${\cal T}_{(p;q),[i]} ^{i_1 \cdots i_p}  {\cal R}^{[i]}$
is already given by
(\ref{transKillingtesor}).
Substituting these results into
(\ref{variKpotSUNplus1tensor})
we reach
(\ref{KllingpotM2}).

On the other hand, from
(\ref{KllingpotM})
and
(\ref{KpotSUNplus1dualtensor})
the Killing potential for dual tensor
is given as
\beqa
\!\!\!\!
\begin{array}{rl}
- i {\cal M}_{(\overline{p};q)}
\!\!\!&\!=\!
{\cal K}_{(\overline{p};q),~[i]} {\cal R}^{[i]}
\!=\!
{\cal T}_{(\overline{p};q) i_1 \cdots i_p,~[i]} {\cal R}^{[i]}
{\cal G}_{(\overline{p};q) j_1 \cdots j_p} ^{i_1 \cdots i_p}
\overline{{\cal T}}_{(\overline{p};q)} ^{j_1 \cdots j_p} \\
\\
&\!+\!
{\cal T}_{(\overline{p};q) i_1 \cdots i_p}
{\cal G}_{(\overline{p};q) j_1 \cdots j_p,~[i]}
^{i_1 \cdots i_p} {\cal R}^{[i]}
\overline{{\cal T}}_{(\overline{p};q)} ^{j_1 \cdots j_p}
\!+\!
{\cal T}_{(\overline{p};q) i_1 \cdots i_p}
{\cal G}_{(\overline{p};q) j_1 \cdots j_p} ^{i_1 \cdots i_p}
\overline{{\cal T}}_{(\overline{p};q),~[i]} ^{j_1 \cdots j_p}
{\cal R}^{[i]} ,~([i]\!=\!(i\hat{i})) ,
\end{array}
\label{variKpotSUNplus1dualtensor}
\eeqa
in which the variation of
$\delta {\cal G}_{(\overline{p};q) j_1 \cdots j_p} ^{i_1 \cdots i_p}$
is computed as
\beqa
\begin{array}{rl}
\delta {\cal G}_{(\overline{p};q) j_1 \cdots j_p} ^{i_1 \cdots i_p}
\!\!\!&=\!
-
\left(
{\displaystyle \! 1 \!+\! \frac{q}{2}}
\right) \!
{\displaystyle \frac{1}{p!}}
[\det {\cal X}]^{1 \!+\! \frac{q}{2}}
\mbox{tr}
\left\{ \!
{\cal X} \!
\left(
\delta {\cal Q} \overline{{\cal Q}}
\!+\!
{\cal Q} \delta \overline{{\cal Q}}
\right) \!
\right\}
[{\cal X}^{-1}]_{~~j_1} ^{i_1}
\cdots
[{\cal X}^{-1}]_{~~j_p} ^{i_p} \\
\\[-8pt]
&
+
{\displaystyle \frac{1}{p!}}
[\det {\cal X}]^{1 \!+\! \frac{q}{2}} \!
\sum_{r\!=\!1} ^p
[{\cal X}^{-1}]_{~~i_1} ^{j_1}
\!\cdots\!
\left( \!
\delta {\cal Q} \overline{{\cal Q}}
\!+\!
{\cal Q} \delta \overline{{\cal Q}} \!
\right)_{~~j_r} ^{i_r}
\!\cdots\!
[{\cal X}^{-1}]_{~~j_p} ^{i_p} .
\end{array}
\label{delGSUNplus1tensor2}
\eeqa
together with
$
\delta [{\cal X}^{-1}]_{~i}^j
\!=\!
\left( 
\delta {\cal Q} \overline{{\cal Q}}
\!+\!
{\cal Q} \delta \overline{{\cal Q}} 
\right)_{~i}^j
$.
Picking up also only the term
$\delta {\cal Q}$
in
(\ref{delGSUNplus1tensor2}),
the following contraction is also easily executed
in the way parallel to the one made in
(\ref{sumdelGSUNplus1tensor}):
\beqa
\!\!\!\!\!\!\!\!
\begin{array}{rl}
{\cal G}_{(\overline{p};q) j_1 \cdots j_p,~\hat{i}i}
^{i_1 \cdots i_p}
\delta {\cal Q}_{i\hat{i}}
&\!=\!
-
\left(
{\displaystyle \! 1 \!+\! \frac{q}{2}}
\right) \!
{\displaystyle \frac{1}{p!}}
[\det {\cal X}]^{1 \!+\! \frac{q}{2}} \!
\mbox{tr} \left( {\cal R}_T \!-\! \Delta \right) \!
[{\cal X}^{-1}]_{~~j_1} ^{i_1}
\cdots
[{\cal X}^{-1}]_{~~j_p} ^{i_p} \\
\\[-8pt]
&+
{\displaystyle \frac{1}{p!}}
[\det {\cal X}]^{1 \!+\! \frac{q}{2}} \!
\sum_{r\!=\!1} ^p
[{\cal X}^{-1}]_{~~i_1} ^{j_1}
\cdots
\left\{ \!
\left( {\cal R}_T \!-\! \Delta \right)_{~~i} ^{i_r}
[{\cal X}^{-1}]_{~j_r} ^{i} \!
\right\}
\cdots
[{\cal X}^{-1}]_{~~j_p} ^{i_p} .
\end{array}
\label{sumdelGSUNplus1dualtensor}
\eeqa
where we have used
$
\delta {\cal Q} \overline{{\cal Q}} {\cal X}
\!=\!
{\cal R}_T \!-\! \Delta
$
again.
$
\overline{{\cal T}}_{(\overline{p};q),~[i]} ^{j_1 \cdots j_p}
{\cal R}^{[i]}
\!=\!
0
$
is also clear
and
${\cal T}_{(\overline{p};q) i_1 \cdots i_p,~[i]} {\cal R}^{[i]}$
is already contributed by
(\ref{transKillingbartesor}).
Putting these results into
(\ref{variKpotSUNplus1dualtensor})
finally we get
(\ref{KllingpotM3}).


\newpage

\setcounter{equation}{0}
\renewcommand{\theequation}{\arabic{section}.\arabic{equation}}

\section{Solution for anomaly-free
$\frac{SO({\bf 10}\!+\!2)}{SU({\bf 5}\!+\!1) \times U(1)}$ supersymmetric $\sigma$-model}

~~~
The invariant K\"{a}hler potentials for a tensor and a dual tensor
are given by
(\ref{KpotSUNplus1tensor})
and
(\ref{KpotSUNplus1dualtensor}),
repectively.
According to
the formula for $A_{\pm }(Y^k;N\!+\!1)$-anomaly defined in the previous section,
for all $N\!+\!1$
it is sufficient to consider only the positive chirality spinor rep
in which all the tensors have an even number of indices.
The anti-symmetric tensor with rank 2 is identified with
the coordinate ${\cal Q}^{ij}$ of the present coset
the $U(1)$ charge of which is 4.
The lowest $p$ and $q$ are
0 and 4 for ${\cal K}_{(p;q)}$
and
1 and $- 4$ for ${\cal K}_{(\overline{p};q)}$.
Then the K\"{a}hler potential in the present anomaly-free
$\frac{SO({\bf 10} \!+\! 2)}{SU({\bf 5} \!+\! 1) \times U({\bf 1})}$
SUSY $\sigma$-model is given by
\beqa
\begin{array}{rl}
{\cal K}(Z, \overline{Z})
&\!=\!
{\displaystyle \frac{1}{2}}
{\cal K}_{\!f\sigma } ({\cal Q}_f, \overline{\cal Q}_f)
\!+\!
{\cal K}_{(0;4)}
\!+\!
{\cal K}_{(\overline{1};-4)} \\
\\[-6pt]
\!\!&\!=\!
{\displaystyle \frac{1}{2f^2}}
\ln \det {\cal X}_f^{-1}
\!+\!
(\det {\cal X}_f)^2 |h|^2
\!+\!
(\det {\cal X}_f)^{-1}
\overline{{\cal T}}_{\!\!(\overline{1};-4)} {\cal X}_f ^{-1}
{\cal T}_{(\overline{1};-4)} ,
\end{array}
\label{KpotSU5plus1tensor}
\eeqa
where the scalar components of the various $SU({\bf 5}\!+\!1)$ and $U(1)$
representations are denoted by
$Z^\alpha \!\!=\!\! ({\cal Q}_f ^{ij},
{\cal T}_{(\overline{1};-4)} \!\!=\!\! k,h),
(k \!\!=\!\! ({\bf k},k_0), k^\dag k \!\!=\!\! 1)$.
In
(\ref{KpotSU5plus1tensor})
a factor $1 / 2$
in the first term of the R.H.S. is included so as to
get the standard normalization of the kinetic in terms of the Goldstone boson fields.
Then using equations
(\ref{KpotSUNplus1tensor}) and (\ref{KllingpotM2}),
and
(\ref{KpotSUNplus1dualtensor}) and (\ref{KllingpotM3}),
the full Killing potential $(N \!=\! 5)$ is explicitly represented as
\beqa
\!\!\!\!\!\!\!\!
\begin{array}{rl}
\!\!&
-i {\cal M}_{\!f} \!
( \! {\cal Q}_f, \overline{\cal Q}_f ; \delta {\cal G} \! )
\!=\!
-i {\displaystyle \frac{1}{2}}
{\cal M}_{\!f\sigma } \! ( \! {\cal Q}_f, \overline{\cal Q}_f ;
\delta {\cal G} \! )
\!-\!
i {\cal M}_{\!f} \! ( \! {\cal Q}_f, \overline{\cal Q}_f ;
\delta {\cal G} \! )_{(0;4)}
\!-\!
i {\cal M}_{\!f} \! ( \! {\cal Q}_f, \overline{\cal Q}_f ;
\delta {\cal G} \! )_{(\overline{1};-4)} \\
\\[-6pt]
&\!=\!
-\mbox{tr} \!
\left[
\Delta_f ( \! {\cal Q}_f, \overline{\cal Q}_f ;
\delta {\cal G} \! ) 
\right] \!
\left( \!
{\displaystyle \frac{1}{2f^2}}
\!-\!
2 {\cal K}_{(0;4)}
\!+\!
{\cal K}_{(\overline{1};-4)} \!
\right) \!
-
e^{f^2 {\cal K}_{\!f\sigma } ({\cal Q}_f, \overline{\cal Q}_f)}
k^\dag
\Delta_f ( \! {\cal Q}_f, \overline{\cal Q}_f ; \delta {\cal G} \! )
{\cal X}_f ^{-1}
k \\
\\[-6pt]
&\!=\!
-\mbox{tr} \!
\left[
\left( \!
f^2 {\cal Q}_f \delta {\cal A} {\cal Q}^\dag _f
\!-\!
\delta {\cal A}^{\mbox{\scriptsize T}}
\!-\!
f \delta {\cal B} {\cal Q}^\dag _f
\!+\!
f {\cal Q}_f \delta {\cal B}^\dag \!
\right) \!
{\cal X}_f
\right] \!
\left( \!
{\displaystyle \frac{1}{2f^2}}
\!-\!
2 {\cal K}_{(0;4)}
\!+\!
{\cal K}_{(\overline{1};-4)} \!
\right) \\
\\[-6pt]
&~-
e^{f^2 {\cal K}_{\!f\sigma } ({\cal Q}_f, \overline{\cal Q}_f)}
k^\dag  \!
\left( \!
f^2 {\cal Q}_f \delta {\cal A} {\cal Q}^\dag _f
\!-\!
\delta {\cal A}^{\mbox{\scriptsize T}}
\!-\!
f \delta {\cal B} {\cal Q}^\dag _f
\!+\!
f {\cal Q}_f \delta {\cal B}^\dag \!
\right) \!
{\cal X}_f
{\cal X}_f ^{-1}
k,
\end{array}
\label{KillingpotSU5plus1}
\eeqa
where we have used
(\ref{RRTChif}),
(\ref{formKillingpotMf})
and
$
(\det {\cal X}_f)^{-1}
\!=\!
e^{f^2 {\cal K}_{\!f\sigma }}
$.
Comparing
(\ref{KillingpotSU5plus1})
with the expression for the Killing potential
(\ref{KillingpotM})
we obtain
a $f$-deformed Killing potential ${\cal M}_{\!f \sigma }~(N \!=\! 5)$
\beqa
\left.
\begin{array}{rl}
-i{\cal M}_{\!f \sigma \delta {\cal B}}
\!\!&\!=\!
-f {\cal X}_f {\cal Q}_f \!
\left( \!
{\displaystyle \frac{1}{2f^2}}
\!-\!
2 {\cal K}_{(0;4)}
\!+\!
{\cal K}_{(\overline{1};-4)} \!
\right) \!
-
f e^{f^2 {\cal K}_{\!f\sigma } ({\cal Q}_f, \overline{\cal Q}_f)}
k k^\dag {\cal Q}_f , \\
\\[-6pt]
-i{\cal M}_{\!f \sigma \delta {\cal B}^\dag }
\!\!&\!=\!
f {\cal Q}^\dag _f {\cal X}_f \!
\left( \!
{\displaystyle \frac{1}{2f^2}}
\!-\!
2 {\cal K}_{(0;4)}
\!+\!
{\cal K}_{(\overline{1};-4)} \!
\right) \!
+
f e^{f^2 {\cal K}_{\!f\sigma } ({\cal Q}_f, \overline{\cal Q}_f)}
{\cal Q}_f ^\dag k k^\dag , \\
\\[-6pt]
-i{\cal M}_{\!f \sigma \delta {\cal A}}
\!\!&\!=\!
\left( \!
1_{N+1} \!-\! 2 f^2 {\cal Q}^\dag _f{\cal X}_f {\cal Q}_f \!
\right) \!\!
\left( \!
{\displaystyle \frac{1}{2f^2}}
\!-\!
2 {\cal K}_{(0;4)}
\!+\!
{\cal K}_{(\overline{1};-4)} \!
\right) \\
\\[-6pt]
&~~~~~~~~~~~~~~~~~~~~~~~~~~~~~~~+
e^{f^2 {\cal K}_{\!f\sigma } ({\cal Q}_f, \overline{\cal Q}_f)} \!
\left( \!
\overline{k} k^{\mbox{\scriptsize T}}
\!-\!
f^2 {\cal Q}_f ^\dag k k^\dag {\cal Q}_f \!
\right) .
\end{array}
\right\}
\label{SO12U6KillingpotMf}
\eeqa
Substituting 
(\ref{RRTChif}) and (\ref{inverse1plusQQf})
into the last equation of
(\ref{SO12U6KillingpotMf})
and using again the auxiliary function
$
\lambda_f
\!=\!
r_{\!f} r_{\!f} ^\dag
\!-\!
f^2 q \overline{r}_{\!f} r_{\!f} ^{\mbox{\scriptsize T}}q^\dag
$,
we get the $f$-deformed Killing potential
${\cal M}_{\!f \sigma \delta {\cal A}}$
as

\beqa
\!\!\!\!\!\!\!\!\!\!
\BA{ll}
&~
-i{\cal M}_{\!f \sigma \delta {\cal A}}
= \\
\\[-8pt]
&
\left[ \!\!\!\!\!
\BA{cc}
\BA{c}
\left\{ \!
1_N \!-\! 2q^\dag \chi_{\!f} q
\!+\!
2
{\displaystyle \frac{Z^2}{f^2}} \!
\left( \!
f^2 q^\dag \chi_{\!f} \lambda_{\!f} \chi_{\!f} q
\!+\!
f^2 q^\dag \chi_{\!f} q \overline{r}_{\!f}
r^{\mbox{\scriptsize T}}_{\!f}
\right.
\right. \\
\left.
\left.
+
f^2 \overline{r}_{\!f} r^{\mbox{\scriptsize T}}_{\!f}
q^\dag \chi_{\!f} q
\!-\!
\overline{r}_{\!f} r^{\mbox{\scriptsize T}}_{\!f}
\right)^{} \!
\right\} \\
\times
\left( \!
{\displaystyle \frac{1}{2f^2}}
\!-\!
2 {\cal K}_{(0;4)}
\!+\!
{\cal K}_{(\overline{1};-4)} \!
\right) \\
\\[-10pt]
+
e^{f^2 {\cal K}_{\!f\sigma }} \!\!
\left\{ \!
\overline{{\bf k}} {\bf k}^{\mbox{\scriptsize T}}
\!-\!
(f q^\dag {\bf k} \!-\! \overline{r}_{\!f} k_0)
(f {\bf k}^\dag q \!-\! \overline{k}_0 r^{\mbox{\scriptsize T}}_{\!f}) \!
\right\}
\EA
&
\BA{c}
-
2 {\displaystyle \frac{1}{f}} q^\dag \chi_{\!f} r_{\!f}
\!+\! 
2 {\displaystyle \frac{Z^2}{f}} \!
\left(
q^\dag \chi_{\!f} \lambda_{\!f} \chi_f r_{\!f}
\right. \\
\left.
+
\overline{r}_{\!f} r^{\mbox{\scriptsize T}}_{\!f}
q^\dag \chi_{\!f} r_{\!f}
\right) \\
\times
\left( \!
{\displaystyle \frac{1}{2f^2}}
\!-\!
2 {\cal K}_{(0;4)}
\!+\!
{\cal K}_{(\overline{1};-4)} \!
\right) \\
\\[-10pt]
+
e^{f^2 {\cal K}_{\!f\sigma }} \!\!
\left\{ \!
\overline{{\bf k}} k_0
\!-\!
(f q^\dag {\bf k} \!-\! \overline{r}_{\!f} k_0)
{\bf k}^\dag r_{\!f} \!
\right\}
\EA \\
\\[-6pt]
\BA{c}
- 2 {\displaystyle \frac{1}{f}} r^\dag_{\!f} \chi_{\!f} q
\!+\!
2 {\displaystyle \frac{Z^2}{f}} \!
\left( \!
r^\dag_{\!f} \chi_{\!f} \lambda_{\!f} \chi_{\!f} q
+
r^\dag_{\!f} \chi_{\!f} q \overline{r}_{\!f}
r^{\mbox{\scriptsize T}}_{\!f} \!
\right) \\
\times
\left( \!
{\displaystyle \frac{1}{2f^2}}
\!-\!
2 {\cal K}_{(0;4)}
\!+\!
{\cal K}_{(\overline{1};-4)} \!
\right) \\
\\[-10pt]
+
e^{f^2 {\cal K}_{\!f\sigma }} \!\!
\left\{ \!
\overline{k}_0 {\bf k}^{\mbox{\scriptsize T}}
\!-\!
r^\dag_{\!f} {\bf k}
(f {\bf k}^\dag q \!-\! \overline{k}_0 r^{\mbox{\scriptsize T}}_{\!f}) \!
\right\}
\EA
&\!\!\!\!
\BA{c}
1
\!-\!
2 {\displaystyle \frac{1}{f^2}} r^\dag_{\!f} \chi_{\!f} r_{\!f}
\!+\!
2{\displaystyle \frac{Z^2}{f^2}}
r^\dag_{\!f} \chi_{\!f} \lambda_{\!f} \chi_f r_{\!f} \\
\times
\left( \!
{\displaystyle \frac{1}{2f^2}}
\!-\!
2 {\cal K}_{(0;4)}
\!+\!
{\cal K}_{(\overline{1};-4)} \!
\right) \\
\\[-10pt]
+
e^{f^2 {\cal K}_{\!f\sigma }} \!\!
\left( \!
\overline{k}_0 k_0 
\!-\!
r^\dag_{\!f} {\bf k} {\bf k}^\dag r_{\!f} \!
\right)
\EA
\EA  \!\!\!\!
\right] \!
\EA \!\! .
\label{SO12U6KillingpotAf}
\eeqa\\[-16pt]
Using the relations
(\ref{relation1f})
and
(\ref{relation2f}),
we get a more
compact form of
$-i {\cal M}_{\!f \sigma \delta {\cal A}}$
as\\[-18pt]
\beqa
\!\!\!\!\!\!\!\!\!\!
\BA{ll}
&~
-i{\cal M}_{\!f \sigma \delta {\cal A}}
= \\
\\[-8pt]
&
\left[ \!\!\!\!\!\!
\BA{cc}
\BA{c}
\left\{ \!
1_N \!-\! 2q^\dag \chi_{\!f} q
\!+\!
2
{\displaystyle \frac{Z^2}{f}} \!\!
\left( \!\!
f q^\dag \chi_{\!f} r_{\!f} r^\dag_{\!f} \chi_{\!f} q
\!-\!
{\displaystyle \frac{1}{f}}
\overline{\chi }_{\!f} \overline{r}_{\!f}
r^{\mbox{\scriptsize T}}_{\!f} \overline{\chi }_{\!f} \!\!
\right) \!\!
\right\} \\
\times
\left( \!
{\displaystyle \frac{1}{2f^2}}
\!-\!
2 {\cal K}_{(0;4)}
\!+\!
{\cal K}_{(\overline{1};-4)} \!
\right) \\
\\[-10pt]
+
e^{f^2 {\cal K}_{\!f\sigma }} \!\!
\left\{ \!
\overline{{\bf k}} {\bf k}^{\mbox{\scriptsize T}}
\!-\!
(f q^\dag {\bf k} \!-\! \overline{r}_{\!f} k_0)
(f {\bf k}^\dag q \!-\! \overline{k}_0 r^{\mbox{\scriptsize T}}_{\!f}) \!
\right\}
\EA
&
\BA{c}
- 2
{\displaystyle \frac{Z^2}{f}} q^\dag \chi_{\!f} r_{\!f} \\
\times
\left( \!
{\displaystyle \frac{1}{2f^2}}
\!-\!
2 {\cal K}_{(0;4)}
\!+\!
{\cal K}_{(\overline{1};-4)} \!
\right) \\
\\[-10pt]
+
e^{f^2 {\cal K}_{\!f\sigma }} \!\!
\left\{ \!
\overline{{\bf k}} k_0
\!-\!
(f q^\dag {\bf k} \!-\! \overline{r}_{\!f} k_0)
{\bf k}^\dag r_{\!f} \!
\right\}
\EA \\
\\[-6pt]
\BA{c}
- 2 {\displaystyle \frac{Z^2}{f}}
r^\dag_{\!f} \chi_{\!f} q \\
\times
\left( \!
{\displaystyle \frac{1}{2f^2}}
\!-\!
2 {\cal K}_{(0;4)}
\!+\!
{\cal K}_{(\overline{1};-4)} \!
\right) \\
\\[-10pt]
+
e^{f^2 {\cal K}_{\!f\sigma }} \!\!
\left\{ \!
\overline{k}_0 {\bf k}^{\mbox{\scriptsize T}}
\!-\!
r^\dag_{\!f} {\bf k}
(f {\bf k}^\dag q \!-\! \overline{k}_0 r^{\mbox{\scriptsize T}}_{\!f}) \!
\right\}
\EA
&\!\!\!\!
\BA{c}
\left\{ \!
{\displaystyle \frac{1}{f^2}
(2Z^2 \!-\! 1) \!+\! 1 \!-\! \frac{1}{f^2}} \!
\right\} \\
\times
\left( \!
{\displaystyle \frac{1}{2f^2}}
\!-\!
2 {\cal K}_{(0;4)}
\!+\!
{\cal K}_{(\overline{1};-4)} \!
\right) \\
\\[-10pt]
+
e^{f^2 {\cal K}_{\!f\sigma }} \!\!
\left( \!
\overline{k}_0 k_0
\!-\!
r^\dag_{\!f} {\bf k} {\bf k}^\dag r_{\!f} \!
\right)
\EA
\EA  \!\!\!\!\!\!
\right] \!
\EA \!\! .
\label{SO12U6KillingpotAf2}
\eeqa\\[-16pt]
The K\"{a}hler potentials are
given as
$
{\cal K}_{(0; 4)}
\!=\!
(\det {\cal X}_{f})^2 |h|^2
$
and
$
{\cal K}_{(\overline{1};- 4)}
\!=\!
(\det {\cal X}_{f})^{-1}
k^\dag {\cal X}_{f} ^{-1} k
$.
The $f$-deformed reduced scalar potential is given by
(\ref{specialscalerpotf})
in which each $f$-deformed Killing potential is computed
straightforwardly as\\[-20pt]
\beqa
\BA{rl}
\mbox{tr}
\left(
-i{\cal M}_{\!f \sigma \delta {\cal A}}
\right)
\!\!&\!=\!
\left\{ \!
\left( \!
1 \!-\! 2 {\displaystyle \frac{1}{f^2}} \!
\right) \!
N
\!+\!
2 {\displaystyle \frac{1}{f^2}} \mbox{{\bf tr}}(\chi_{\!f})
\!+\!
2 {\displaystyle \frac{Z^2}{f^2}}
\mbox{{\bf tr}}(\chi_{\!f} r_{\!f} r^\dag_{\!f})
\!-\!
4 {\displaystyle \frac{Z^2}{f^2}}  
\mbox{{\bf tr}}(\chi_{\!f} r_{\!f} r^\dag_{\!f} \chi_{\!f}) 
\right. \\
\\[-14pt]
&
\left.
+
{\displaystyle \frac{1}{f^2}} (2 Z^2 \!-\! 1)
\!+\! 
1 \!-\! {\displaystyle \frac{1}{f^2}}
\right\} \!\!
\left( \!
{\displaystyle \frac{1}{2f^2}}
\!-\!
2 {\cal K}_{(0;4)}
\!+\!
{\cal K}_{(\overline{1};-4)} \!
\right) \\
\\[-10pt]
&
+
e^{f^2 {\cal K}_{\!f\sigma }}
\mbox{{\bf tr}} \!
\left\{ \!
\overline{{\bf k}} {\bf k}^{\mbox{\scriptsize T}}
\!-\!
(f q^\dag {\bf k} \!-\! \overline{r}_{\!f} k_0)
(f {\bf k}^\dag q \!-\! \overline{k}_0 r^{\mbox{\scriptsize T}}_{\!f}) \!
\right\}
\!+\!
e^{f^2 {\cal K}_{\!f\sigma }} \!\!
\left( \!
\overline{k}_0 k_0
\!-\!
r^\dag_{\!f} {\bf k} {\bf k}^\dag r_{\!f} \!
\right) . 
\EA
\label{SO12U6trM2f1}
\eeqa\\[-16pt]
Using approximations for
$\mbox{{\bf tr}}(\!\chi_{\!f}\!)$,
$\mbox{{\bf tr}}(\!\chi_{\!f} r_{\!f} r^\dag_{\!f}\!)$ etc.
and
(\ref{rdagchichirftrrrdagf}),
$
\mbox{tr} \!
\left(\!
-i\!{\cal M}_{\!f \sigma \delta {\cal A}}\!
\right)
$
is  calculated as\\[-16pt]
\beqa
\BA{rl}
\mbox{tr}
\left(
-i{\cal M}_{\!f \sigma \delta {\cal A}}
\right)
\!\!&
\!=\!
\left\{ \!
1
\!+\!
\left( \!
1 \!-\! 2 {\displaystyle \frac{1}{f^2}} \!
\right) \!
N
\!+\!
2 {\displaystyle \frac{1}{f^2}} (2Z^2 \!-\! 1) \!
< \!\!\chi_{\!f}\!\! > \!
\right\} \!\!
\left( \!
{\displaystyle \frac{1}{2f^2}}
\!-\!
2 {\cal K}_{(0;4)}
\!+\!
{\cal K}_{(\overline{1};-4)} \!
\right) \\
\\[-10pt]
&
+
e^{f^2 {\cal K}_{\!f\sigma }}
\mbox{{\bf tr}} \!
\left\{ \!
\overline{{\bf k}} {\bf k}^{\mbox{\scriptsize T}}
\!-\!
(f q^\dag {\bf k} \!-\! \overline{r}_{\!f} k_0)
(f {\bf k}^\dag q \!-\! \overline{k}_0 r^{\mbox{\scriptsize T}}_{\!f}) \!
\right\}
\!+\!
e^{f^2 {\cal K}_{\!f\sigma }} \!
\left( \!
\overline{k}_0 k_0 
\!-\!
r^\dag_{\!f} {\bf k} {\bf k}^\dag r_{\!f} \!
\right) \! ,
\EA
\label{approxSO12U6trM2f1}
\eeqa\\[-16pt]
$
\mbox{tr} \!
\left(\!
-i\!{\cal M}_{\!f \sigma \delta {\cal A}}\!
\right)^2
$
is also computed,
though we omit the result
since its expression is very lengthy.
Substituting
(\ref{approxSO12U6trM2f1})
and
the
$
\mbox{tr} \!
\left(\!
-i\!{\cal M}_{\!f \sigma \delta {\cal A}}\!
\right)^2
$
into
(\ref{specialscalerpotf}),
we get the $f$-deformed reduced scalar potential
$V_{\!f{\mbox{\scriptsize redSC}}}$
as\\[-16pt]
\beqa
\!\!\!\!\!\!
\BA{rl}
&\!\!\!\!
V_{\!f{\mbox{\scriptsize redSC}}}
\!=\!
{\displaystyle
\frac{g^2 _{U(1)}}{2(N \!+\! 1)}
}
\!\cdot\!
\left[
\xi
\!+\!\!
\left\{ \!
1
\!\!+\!\!
\left( \!\!
1 \!\!-\!\! 2 {\displaystyle \frac{1}{f^2}} \!\!
\right) \!\!
N
\!\!+\!\!
2 {\displaystyle \frac{1}{f^2}} (\!2Z^2 \!\!-\!\! 1\!) \!\!
< \!\!\chi_{\!f}\!\! > \!\!
\right\} \!\!
\left( \!
{\displaystyle \frac{1}{2f^2}}
\!-\!
2 {\cal K}_{(0;4)}
\!+\!
{\cal K}_{(\overline{1};-4)} \!
\right) 
\right. \\
\\[-12pt]
&
\left.
\!+\!
e^{^{^{f^2 {\cal K}_{\!f\sigma }}}} \!\!
\left\{ \!
1
\!+\!
N |{\bf k}|^2 \!
\!-\!
\left( \!
{\displaystyle \frac{1 \!-\! Z^2}{Z^2}} \!+\! N |{\bf k}|^2 \!
\right) \!
{\displaystyle \frac{1}{< \!\!\chi_{\!f}\!\! >}} \!
\right\}
\right]^2
\!\!+\!
2
{\displaystyle
\frac{g^2 _{SU(N \!+\! 1)}}{N \!+\! 1}
}
\!\cdot\!
\left[
{\displaystyle \frac{{N \!+\! 1}^{^{^{.}}}}{4}} \!
\left[
1 \!+\! N \!-\! 4 {\displaystyle \frac{1}{f^2}} \!
\left( \!
1 \!-\! {\displaystyle \frac{1}{f^2}} \!
\right) \! N 
\right.
\right. \\
\\[-14pt]
&
\left.
-
4 {\displaystyle \frac{1}{f^2}} \!
\left\{ \!
{\displaystyle \frac{1}{f^2}}(2 Z^2 \!\!-\!\! 1)
\!-\!
\left( \!\!
1 \!\!-\!\! {\displaystyle \frac{1}{f^2}} \!\!
\right) \!\!
Z^2 \!
\right\} \!\!
< \!\!\chi_{\!f}\!\! >
\!+\! 
4 {\displaystyle \frac{1}{f^4}}(2 Z^4 \!\!-\!\! 1) \!
< \!\chi_{\!f}\! >^2
\right] \!\!
\left( \!
{\displaystyle \frac{1}{2f^2}}
\!-\!
2 {\cal K}_{(0;4)}
\!+\!
{\cal K}_{(\overline{1};-4)} \!
\right)^{\!\!2} \\
\\[-10pt]
&
+\!\!
\left[ \!
2 e^{^{^{f^2 {\cal K}_{\!f\sigma }}}} \!
\left\{ \!
\left(\!\!1 \!\!-\!\! 2 {\displaystyle \frac{1}{f^2}} \!\!\right)
\!\! N
\!\!+\!\!
2 {\displaystyle \frac{1}{f^2}} (1 \!\!-\!\! Z^2)
\!\!+\!\!
2 {\displaystyle \frac{1}{f^2}} (2 Z^2 \!\!-\!\! 1) \!\!
< \!\!\chi_{\!f}\!\! > \!
\right\}
\right.
\\
\\[-14pt]
&~~~~
\!\times\!
\left\{ \!
(N \!\!+\!\! 1)|{\bf k}|^2
\!\!-\!\!
\left( \!
{\displaystyle \frac{1 \!-\! Z^2}{Z^2}}
(1 \!\!-\!\! |{\bf k}|^2) \!\!+\!\! N |{\bf k}|^2 \!
\right) \!
{\displaystyle \frac{1}{< \!\!\chi_{\!f}\!\! >}} \!
\right\}
-
8 {\displaystyle \frac{1}{f^2}}
(1 \!-\! Z^2) |{\bf k}|^2
e^{f^2 {\cal K}_{\!f\sigma }} \!
\left( \!
1
\!-\!
N {\displaystyle \frac{1}{< \!\!\chi_{\!f}\!\! >}} \!
\right) \\
\\[-10pt]
&
\left.
\!+\!
2 e^{f^2 {\cal K}_{\!f\sigma }} \!\!
\left\{ \!
{\displaystyle \frac{1}{f^2}
(2Z^2 \!-\! 1) \!+\! 1 \!-\! \frac{1}{f^2}} \!
\right\} \!\!
\left( \!
1 \!-\! |{\bf k}|^2
\!-\!
|{\bf k}|^2
{\displaystyle \frac{1 \!-\! Z^2}{Z^2}}
{\displaystyle \frac{1}{< \!\!\chi_{\!f}\!\! >}}
\right) \!
\right] \!
\left( \!
{\displaystyle \frac{1}{2f^2}}
\!-\!
2 {\cal K}_{(0;4)}
\!+\!
{\cal K}_{(\overline{1};-4)} \!
\right) \\
\\[-8pt]
&
+
e^{2 f^2 {\cal K}_{\!f\sigma }} \!\!
\left[ \!
1 \!\!+\!\! N \! (N \!\!+\!\! 2) \! |{\bf k}|^4
\!\!-\!\!
2 N \!|{\bf k}|^2 \!\!
\left\{ \!\!
{\displaystyle \frac{1 \!\!-\!\! Z^2}{Z^2}}
\!+\!\! (N \!\!+\!\! 1) |{\bf k}|^2 \!
\right\} \!\!
{\displaystyle \frac{1}{< \!\!\chi_{\!f}\!\! >}}
\!+\!\!
N |{\bf k}|^2 \!\!
\left\{ \!\!
{\displaystyle \frac{1 \!\!-\!\! Z^4}{Z^4}} \!+\!\! N |{\bf k}|^2 \!
\right\} \!\!
{\displaystyle \frac{1}{< \!\!\chi_{\!f}\!\! >^2}} \!
\right] \\
\\[-10pt]
&
- {\displaystyle \frac{1}{4}} \!
\left[ \!
\left\{
1
\!\!+\!\!
\left( \!\!
1 \!\!-\!\! 2 {\displaystyle \frac{1}{f^2}} \!\!
\right) \!\!
N
\!\!+\!\!
2 {\displaystyle \frac{1}{f^2}} (\!2Z^2 \!\!-\!\! 1\!) \!\!
< \!\!\chi_{\!f}\!\! > \!\!
\right\} \!\!
\left( \!\!
{\displaystyle \frac{1}{2f^2}}
\!\!-\!\!
2 {\cal K}_{(0;4)}
\!\!+\!\!
{\cal K}_{(\overline{1};-4)} \!\!
\right)^{^{.}} \!\!
\right.\\
\\[-14pt]
&
\left.
\left.
~~~~~~~~~~~~~~~~~~~~~~~~~~~~~~~~~~~~~~~~~~
\!+\!
e^{f^2 {\cal K}_{\!f\sigma }} \!
\left\{ \!
1
\!\!+\!\!
N \! |{\bf k}|^2 \!
\!-\!\!
\left( \!
{\displaystyle \frac{1 \!-\! Z^2}{Z^2}} \!+\!\! N \! |{\bf k}|^2 \!
\right) \!\!
{\displaystyle \frac{1}{< \!\!\chi_{\!f}\!\! >}} \!
\right\}
\right]^{2}
\right]~(N \!=\! 5) .
\EA
\label{specialscalerpot2f2final}
\eeqa\\[-8pt]
In the above we have approximated the terms
$
\mbox{{\bf tr}}\!\left(\overline{r}_{\!f} k^\dag q \right),
r^{\mbox{\scriptsize T}}_{\!f} q^\dag k,
\mbox{{\bf tr}}\!
\left(q^\dag k r^{\mbox{\scriptsize T}}_{\!f} \right)~
\mbox{and}~
k^\dag q \overline{r}_{\!f}
$
to be zero.
We have also used the relation
$|{\bf k}|^2 \!+\! |k_0 \!|^2 \!=\! 1$.
Then we have only magnitude of the vector ${\bf k}$, $|{\bf k}|^2$ and
magnitude of $k_0$, $|k_0|^2$.
This means that the $f$-deformed reduced scalar potential is
manifestly invariant under an $SU({\bf 5})$ and a $U(1)$ transformations,
respectively.
Variations of
(\ref{specialscalerpot2f2final})
with respect to $Z^2$ and $< \!\!\chi_{\!f}\!\! >$
lead to the following cubic equation for
$
\left(
1 / 2 f^2
\!-\!
2 {\cal K}_{(0;4)}
\!+\!
{\cal K}_{(\overline{1};-4)}
\right)(\!\equiv\! E)
$:\\[-14pt]
\beqa
\BA{rl}
&
8 (N \!+\! 1)
{\displaystyle \frac{1}{f^6}} \!
< \!\!\chi_{\!f}\!\! > \!\!
\left\{
\left( \!
1 \!-\! Z^2
\right) \!
< \!\!\chi_{\!f}\!\! > \!
-
{\displaystyle \frac{1}{4}} f^2 \!
\left( \!
1 \!-\!
{\displaystyle \frac{1}{f^2}} \!
\right) \!
\right\} \!
E^3 \\
\\[-16pt]
&
\!+
{\displaystyle \frac{1}{f^4}} \!
{\displaystyle \frac{1}{Z^{\!4}}}
{\displaystyle \frac{1}{< \!\!\chi_{\!f}\!\! >}}
e^{f^2 {\cal K}_{\!f\sigma }} \!\!
\left[{}^{^{^{^{^{^{^{.}}}}}}} \!\!\!
\!\!-\!
\left( \! N \!+\! 1 \! \right)
\!-\!
8
\!+\!
4 N \!
\left( \! 2 \!-\! f^2 \! \right)
\!+\! 4 \!
\left\{ \!
4 \!-\! f^2
\!-\!
\! N \!
\left( \! 2 \!-\! f^2 \! \right) \!
\right\} \!
|{\bf k}|^2
\right. \\
\\[-14pt]
&
\!+
\left[
43
\!-\!\!
f^2
\!-\!\!
N \! \left( 13 \!-\! 7 \! f^2 \right)
\!+\!
\left\{
\!-\! 48 \!+\! 5 N \!\!+\!\! 13 N^2
\!+\!
\left( 8 \!\!-\!\! 7 N \!\!-\!\! 7 N^2 \right) \!
f^2
\right\} \!
|{\bf k}|^2
\right] \!
Z^4 \!
\!-\!
32
\left\{ \!
1 \!\!+\!\!
\left( N \!-\! 2 \right) \! |{\bf k}|^2
\right\} \!
Z^6 \\
\\[-16pt]
&
\left.
\!+\!
2 \!
\left[
N \!\!+\!\! 1
\!\!+\!\!
4 \! \left( \! 1 \!\!-\!\! |{\bf k}|^2 \! \right)
\!-\!
8 \! \left( \! 1 \!\!-\!\! |{\bf k}|^2 \! \right) \!
Z^2
\!\!-\!\!
4 \!
\left\{ \!
3 \!\!-\!\!
\left( \! 3 N \! \!+\! 2 \! \right) \! |{\bf k}|^2 \!
\right\} \!\!
Z^4
\!\!+\!\!
2 \!
\left\{ \!
11 \!\!-\!\! N
\!\!+\!\!
\left( \! N^2 \!\!-\!\! 11 N \!\!-\!\! 8 \! \right) \!
|{\bf k}|^2
\right\} \!\!
Z^6
\right] \!
< \!\!\chi_{\!f}\!\! > \! 
{}^{^{^{^{^{^{^{.}}}}}}} \!\!\!\!
\right] \!\!
E^2 \\
\\[-20pt]
&
\!+ 2
{\displaystyle \frac{1}{f^2}}
{\displaystyle \frac{1}{Z^{\!6}}} \!
{\displaystyle \frac{1}{< \!\!\chi_{\!f}\!\! >^{\!3}}}
e^{2 f^2 {\cal K}_{\!f\sigma }} \!\!\! 
\left[{}^{^{^{^{^{^{^{.}}}}}}} \!\!\!\!\!
\left[{}^{^{^{^{.}}}} \!\!\!
2
\!-\!
2 \!
\left\{
2 
\!-\!
N \!
\left(
2 \!-\! f^2
\right)
\right\} \!
|{\bf k}|^2
\!-\!
N \!\!
\left\{ \!
4
\!-\!\!
f^2
\!\!-\!\!
N \!\! \left( 2 \!-\!\! f^2 \right)
\right\} \!
|{\bf k}|^4
{}^{^{^{^{.}}}} \!\!\! \right] \!\!
Z^2 
\right. \\
\\[-12pt]
&
- 2 \!
\left\{
2
\!+\! 3 \!
\left( N \!-\!\! 2 \right) \! |{\bf k}|^2
\right\} \!
Z^4
\!+\! 2 \!
\left( 1 \!\!-\!\! N |{\bf k}|^2 \right) \!
\left\{
1
\!+\!\!
\left( N \!-\!\! 2 \right) \! |{\bf k}|^2
\right\} \!
Z^6 \\
\\[-10pt]
&
- 2 \!
\left[{}^{^{^{^{.}}}} \!\!\!
1 \!\!-\!\! |{\bf k}|^2
\!\!-\!\!
N \! |{\bf k}|^2
\!\!-\!\!
\left( \! 1 \!\!-\!\! |{\bf k}|^2 \! \right) \!
\left( \! 1 \!\!-\!\! N  \! |{\bf k}|^2 \! \right) \! Z^2 \!
\!\!-\!\!
\left\{ \!
3 \!-\!
\left( \! 3 N \!\!+\!\! 2 \right) \!
|{\bf k}|^2 \!
\right\} \!
\! Z^4
\!\!+\!\!
\left( \! 1 \!\!-\!\! N  \! |{\bf k}|^2 \! \right) \!\!
\left\{ \!
3 \!-\!
\left( \! N \!\!-\!\! 2 \right) \!
|{\bf k}|^2 \!
\right\} \!\!
Z^6
{}^{^{^{^{.}}}} \!\!\! \right] \!\!
< \!\!\chi_{\!f}\!\! > \\
\\[-16pt]
&
\left.
\!+ 2
|{\bf k}|^2 Z^2 \!
\left\{
2N \!\!+\!\! 1
\!-\!
\left( N \!\!+\!\! 1 \right) \! 
\left( 1 \!\!-\!\! 2 N |{\bf k}|^{\!2} \right)
 Z^{\!4}
\right\} \!
< \!\!\chi_{\!f}\!\! >^2 \!\!\!\!
{}^{^{^{^{^{^{^{.}}}}}}} \!
\right] \!\!
E \\
\\[-18pt]
&
\!+
2 N |{\bf k}|^2
{\displaystyle \frac{1}{Z^4}}
{\displaystyle \frac{1}{< \!\!\chi_{\!f}\!\! >^3}}
e^{3 f^2 {\cal K}_{\!f\sigma }} \!
\left\{ \!
\left(
1 \!-\! N |{\bf k}|^2
\right)
{\displaystyle \frac{1 \!-\! Z^2}{Z^2}}
{\displaystyle \frac{1}{< \!\!\chi_{\!f}\!\! >}} \!
-
|{\bf k}|^2 \!
\right\}
= 0~(N \!=\! 5) .
\EA
\label{VvariZf2chif3}
\eeqa
If only the first term in
(\ref{VvariZf2chif3})
is taken up,
the solutions for
$Z^2$ and $< \!\!\chi_{\!f}\!\! >$
are realized,
which has already been obtained in I
\cite{SJCF.08}.
Instead of such solutions, here we seek for another solutions.
For this aim, the last term is assumed to vanish:
$
\left( \!
1 \!\!-\!\! N |{\bf k}|^2 \!
\right) \!
(1 \!\!-\!\! Z^2) Z^{-2}
\cdot\!
< \!\!\chi_{\!f}\!\! >^{\!-1} \!
- |{\bf k}|^2
\!=\! 0~(N \!\!=\!\! 5) 
$.
Due to this relation,
(\ref{VvariZf2chif3})
becomes a quadratic equation for
$E$.
It is necessary to analyze a value and especially a sign of $E$
which brings
positive definiteness of a matter-extended K\"{a}hler metric $(E \!>\! 0)$
or
negative kinetic-energy ghosts $(E \!<\! 0)$
\cite{NNH.01}.
The $Z^2$ and $< \!\!\chi_{\!f}\!\! >$ are connected with each other
through $|{\bf k}|^2$.
Substituting
$
< \!\!\chi_{\!f}\!\! >
=\!
(1 \!\!-\!\! N |{\bf k}|^2) |{\bf k}|^{-2}
\!\cdot\!
(1 \!\!-\!\! Z^2) Z^{-2}
$
derived from the above relation into
the variational equation for $Z^2$,
we obtain an optimized equation for $Z^2$,
which we omit here since it is very lengthy.
Taking only the zeroth and first order of
$Z^2~(0 \! < \!\! Z^2 \!\! < \!\! 1)$ in
the optimized equation,
we finally reach our ultimate goal of solution for $Z^2~(N \!\!=\!\! 5)$.
Putting this solution into
$
< \!\!\chi_{\!f}\!\! >
=\!
(1 \!\!-\!\! N |{\bf k}|^2) |{\bf k}|^{-2}
\!\cdot\!
(1 \!\!-\!\! Z^2) Z^{-2} ~\!
(N \!\!=\!\! 5)
$,
then we have
\beqa
\!\!\!\!\!\!\!\!\!\!\!\!\!\!
\BA{rl}
&\!\!\!\!
Z^2
\!\!=\!\!
\left[ \!\!
\left[
{}^{^{^{^{^{^{^{.}}}}}}} \!\!\!
g^2 _{U(1)}
\!\cdot\!
\left\{ \!
4
{\displaystyle
\frac{\left( \! 1 \!\!-\!\! N \! |{\bf k}|^2 \! \right)^2}{|{\bf k}|^4}
} \!
E
\!+\!
f^2 e^{f^2 {\cal K}_{\!f\sigma }} \!\!
\right\} \!
\!+\!
2 g^2 _{SU(N \!+\! 1)}
\!\cdot\!
2 \!
\left[ \!
4
e^{^{^{f^2 {\cal K}_{\!f\sigma }}}} \!\!\!
\left( 1 \!\!-\!\! |{\bf k}|^2 \right)
\!-\!
{\displaystyle
\frac{\left( \! 1 \!\!-\!\! N \! |{\bf k}|^2 \! \right)^2}{|{\bf k}|^4}
} \!
E
\!-\!
{\displaystyle \frac{1}{4}} \!
f^2 
e^{f^2 {\cal K}_{\!f\sigma }} \!
{}^{^{^{^{^{^{^{.}}}}}}} \!\!\!
\right] \!\!
{}^{^{^{^{^{^{^{.}}}}}}} \!\!
\right] \!\!
\right] \\
\\
&~
\times\!\!
\left[ \!\!
\left[
{}^{^{^{^{^{^{^{.}}}}}}} \!\!\!
g^2 _{U(1)}
\cdot
\left[ \!
2 {\displaystyle \frac{1 \!-\! N |{\bf k}|^2}{|{\bf k}|^2}} \!
\left\{ \!
f^2 \xi
\!+\!
\left( \!
\left( N \!+\! 1 \right) \! f^2 \!-\! 2 \! N
\!+\!
12
{\displaystyle \frac{1 \!-\! N \! |{\bf k}|^2}{|{\bf k}|^2}} \!
\right) \!
E
\right\}
\!-\!
2 f^2 \!
e^{^{^{f^2 {\cal K}_{f\sigma }}}} \!\!
\left( \!
1
\!+\! N^2 \! |{\bf k}|^2
\!-\!
{\displaystyle \frac{1}{|{\bf k}|^2}} \!
\right)
\right.
\right.
\right. \\
\\
&\!\!\!\!
\left.
~~~~~~~~~~~~~~~~~~~~~~
\!+\!
{\displaystyle \frac{1}{2}}
f^2 e^{f^2 {\cal K}_{f\sigma }} \!
\left\{ \!
f^2 \xi
{\displaystyle \frac{|{\bf k}|^2}{1 \!-\! N |{\bf k}|^2}} \!
{\displaystyle \frac{1}{E}}
\!+\!
8
\!+\!
\left(
\left( N \!+\! 1 \right) \!\! f^2 \!-\! 2 N
\right) \!
{\displaystyle \frac{|{\bf k}|^2}{1 \!-\! N |{\bf k}|^2}} \!
\right\}
\right] \\
\\
&\!\!\!\!
+
2 g^2 _{SU(N \!+\! 1)}
\cdot
{\displaystyle \frac{1}{4}} \!
\left[ \!
{}^{^{^{^{^{^{^{.}}}}}}} \!\!\!
\left\{
\!-\!
2 \!
\left( N \!+\! 1 \right)
\left(
3
\!-\!
f^2
\right)
{\displaystyle \frac{1 \!-\! N |{\bf k}|^2}{|{\bf k}|^2}}
\!+\!
8
\left( N \!+\! 1 \right)
{\displaystyle
\frac{\left( \! 1 \!-\! N |{\bf k}|^2 \! \right)^2}{|{\bf k}|^4}
}
\right\} \!
E
\right. \\
\\
&\!\!\!\!
+
16
e^{^{^{f^2 {\cal K}_{f\sigma }}}} \!
\left[
\left( N \!+\! 1 \right) \! f^2 \!
\left( 1 \!\!-\!\! N \! |{\bf k}|^2 \right) \!
\!+\!
\left( 1 \!\!-\!\! |{\bf k}|^2 \right) \!
\left\{ \!
8
\!-\!
f^2
\!+\!
\left(
2
\!-\!
\left( 2 \!-\! f^2 \right) \!
N
\right)
{\displaystyle \frac{|{\bf k}|^2}{1 \!\!-\!\! N \! |{\bf k}|^2}}
\right\}
\!+\!
\left( 2 \!-\! f^2 \right)
{\displaystyle \frac{|{\bf k}|^2}{1 \!\!-\!\! N \! |{\bf k}|^2}}
\right] \\
\\
&\!\!\!\!
\!+
4 f^4 N |{\bf k}|^2
{\displaystyle \frac{|{\bf k}|^2}{1 \!-\! N |{\bf k}|^2}}
\left( \!
1
\!-\!
{\displaystyle \frac{|{\bf k}|^2}{1 \!-\! N |{\bf k}|^2}} \!
\right)
e^{2 f^2 {\cal K}_{\!f\sigma }}
{\displaystyle \frac{1}{E}} \\
\\
&\!\!\!\!
\!-
4
\left[
{\displaystyle \frac{1 \!-\! N \! |{\bf k}|^2}{|{\bf k}|^2}}
\left\{ \!
\left( N \!+\! 1 \right) \! f^2 \!-\! 2 N
\!+\!
8
{\displaystyle \frac{1 \!-\! N |{\bf k}|^2}{|{\bf k}|^2}}
\right\} \!
E
\!+\!
e^{f^2 {\cal K}_{f\sigma }} \!
\left( \!
1 \!+\! N^2 |{\bf k}|^2
\!-\!
{\displaystyle \frac{1}{|{\bf k}|^2}} \!
\right)
\right] \\
\\
&\!\!\!\!
\left.
\left.
\left.
\!\!-
f^2
e^{f^2 {\cal K}_{f\sigma }} \!
\left[
{\displaystyle \frac{|{\bf k}|^2}{1 \!-\! N |{\bf k}|^2}}
\left\{
\left( N \!+\! 1 \right) \! f^2 \!-\! 2 N
\!+\!
8
{\displaystyle \frac{1 \!-\! N |{\bf k}|^2}{|{\bf k}|^2}}
\right\}
\!-\!
f^2 e^{f^2 {\cal K}_{f\sigma }}
{\displaystyle
\frac{1 \!-\! |{\bf k}|^2
\left( 1 \!+\! N^2 |{\bf k}|^2 \right)}
{1 \!-\! N |{\bf k}|^2}
}
{\displaystyle \frac{1}{E}}
\right] \!\!
{}^{^{^{^{^{^{^{.}}}}}}} \!\!\!
\right] \!
{}^{^{^{^{^{^{^{.}}}}}}} \!\!\!
\right] \!\!
\right]^{\!-1} ,
\EA
\label{solutionforZ2f}
\eeqa\
and
\beqa
\!\!\!\!\!\!\!\!\!\!\!\!\!\!
\BA{rl}
&\!\!\!\!
< \!\! \chi_{\!f} \!\! >
=\!
{\displaystyle \frac{1 \!-\! N |{\bf k}|^2}{|{\bf k}|^2}} \!
\left[ \!\!
\left[
{}^{^{^{^{^{^{.}}}}}} \!\!
g^2 _{U(1)}
\!\cdot\!
\left[
2 {\displaystyle \frac{1 \!-\! N |{\bf k}|^2}{|{\bf k}|^2}}
\left\{ \!
f^2 \xi
\!+\!
\left( \!\!
\left( N \!+\! 1 \right) \! f^2 \!-\! 2 N
\!+\!
10
{\displaystyle \frac{1 \!-\! N |{\bf k}|^2}{|{\bf k}|^2}} \!
\right) \!
E \!
\right\}
\right.
\right.
\right. \\
\\
&\!\!\!\!
\left.
\!-
2 f^2
e^{^{^{f^2 {\cal K}_{\!f\!\sigma }}}} \!
\left( \!
1
\!+\! N^2 |{\bf k}|^2
\!-\!
{\displaystyle \frac{1}{|{\bf k}|^2}} \!
\right)
\!+\!
{\displaystyle \frac{1}{2}}
f^2 e^{f^2 {\cal K}_{f\sigma }} \!
\left\{ \!
f^2 \xi
{\displaystyle \frac{|{\bf k}|^2}{1 \!\!-\!\! N |{\bf k}|^2}}
{\displaystyle \frac{1}{E}}
\!+\!
6
\!+\!
\left(
\left(
N \!+\! 1 \right) \! f^2 \!-\! 2 N 
\right)
{\displaystyle \frac{|{\bf k}|^2}{1 \!\!-\!\! N |{\bf k}|^2}} \!
\right\}
\right] \\
\\
&\!\!\!\!
+
2 g^2 _{SU(N \!+\! 1)}
\cdot
{\displaystyle \frac{1}{4}} \!
\left[ \!\!\!
{}^{^{^{^{^{^{.}}}}}} \!\!
\left\{
\!-\!
2 \!
\left( N \!+\! 1 \right)
\left(
3
\!-\!
f^2
\right)
{\displaystyle \frac{1 \!-\! N |{\bf k}|^2}{|{\bf k}|^2}}
\!+\!
8 N
{\displaystyle
\frac{\left( 1 \!-\! N |{\bf k}|^2 \right)^2}{|{\bf k}|^4}
}
\right\} \!
E
\right. \\
\\
&\!\!\!\!
+
16
e^{^{^{f^2 {\cal K}_{f\sigma }}}} \!\!
\left[
\left( N \!\!+\!\! 1 \right) \! f^2
\left( 1 \!\!-\!\! N |{\bf k}|^2 \right)
\!+\!
\left( 1 \!\!-\!\! |{\bf k}|^2 \right) \!
\left\{
6
\!\!-\!\!
f^2
\!+\!
\left(
2
\!\!-\!\!
\left( 2 \!\!-\!\! f^2 \right) 
N
\right) 
{\displaystyle \frac{|{\bf k}|^2}{1 \!\!-\!\! N |{\bf k}|^2}} \!
\right\}
\!+\!
\left( 2 \!-\! f^2 \right)
{\displaystyle \frac{|{\bf k}|^2}{1 \!\!-\!\! N |{\bf k}|^2}}
\right] \\
\\
&\!\!\!\!
\!-\!
2
f^2
e^{f^2 {\cal K}_{f\sigma }} \!
\!+\!
4 f^4 N |{\bf k}|^2
{\displaystyle \frac{|{\bf k}|^2}{1 \!-\! N |{\bf k}|^2}}
\left( \!
1
\!-\!
{\displaystyle \frac{|{\bf k}|^2}{1 \!-\! N |{\bf k}|^2}} \!
\right)
e^{2 f^2 {\cal K}_{f\sigma }}
{\displaystyle \frac{1}{E}} \\
\\
&\!\!\!\!
\!-
4 \!
\left[ \!
{\displaystyle \frac{1 \!-\! N |{\bf k}|^2}{|{\bf k}|^2}} \!
\left\{ \!
\left( N \!+\! 1 \right) \! f^2 \!-\! 2 N
\!+\!
8
{\displaystyle \frac{1 \!-\! N |{\bf k}|^2}{|{\bf k}|^2}} \!
\right\} \!
E
\!+\!
e^{f^2 {\cal K}_{f\sigma }} \!
\left( \!
1
\!+\! N^2 |k|^2
\!-\!
{\displaystyle \frac{1}{|{\bf k}|^2}} \!
\right)
\right] \\
\\
&\!\!\!\!
\left.
\left.
\left.
\!\!-
f^2 \!
e^{f^2 {\cal K}_{f\sigma }} \!
\left[ \!
{\displaystyle \frac{|{\bf k}|^2}{1 \!-\! N |{\bf k}|^2}} \!
\left\{ \!
\left( N \!+\! 1 \right) \! f^2 \!-\! 2 N
\!+\!
8
{\displaystyle \frac{1 \!-\! N |{\bf k}|^2}{|{\bf k}|^2}} \!
\right\}
\!-\!
f^2 e^{f^2 {\cal K}_{f\sigma }}
{\displaystyle
\frac{1 \!-\! |{\bf k}|^2
\left( 1 \!+\! N^2 |{\bf k}|^2 \right)}
{1 \!-\! N |{\bf k}|^2}
}
{\displaystyle \frac{1}{E}}
\right]
\right]
{}^{^{^{^{^{^{.}}}}}} \!\!
\right] \!\!
\right] \\
\\
&\!\!\!\!
\!\times\!
\left[ \!\!
\left[
{}^{^{^{^{^{^{^{.}}}}}}} \!\!\!
g^2 _{U(1)}
\!\cdot\!
\left\{ \!
4
{\displaystyle
\frac{\left( \! 1 \!\!-\!\! N \! |{\bf k}|^2 \! \right)^2}{|{\bf k}|^4}
} \!
E
\!+\!
f^2 e^{f^2 {\cal K}_{\!f\sigma }} \!\!
\right\}
\!+\!
2 g^2 _{SU(N \!+\! 1)}
\!\cdot\!
2 \!
\left[
4
e^{^{^{f^2 {\cal K}_{\!f\sigma }}}} \!\!\!
\left( 1 \!\!-\!\! |{\bf k}|^2 \right)
\!-\!
{\displaystyle
\frac{\left( \! 1 \!\!-\!\! N \! |{\bf k}|^2 \! \right)^2}{|{\bf k}|^4}
} \!
E
\!-\!
{\displaystyle \frac{1}{4}}
f^2 
e^{f^2 {\cal K}_{\!f\sigma }} \!
{}^{^{^{^{^{^{^{.}}}}}}} \!\!\!
\right] \!
{}^{^{^{^{^{^{^{.}}}}}}} \!\!\!
\right] \!\!
\right]^{\!\!-1} \! .
\EA
\label{solutionforchif}
\eeqa


\newpage

\setcounter{equation}{0}
\renewcommand{\theequation}{\arabic{section}.\arabic{equation}}

\section{Discussions and concluding remarks}

~
By embedding the $SO(2N \!\!+\!\! 1)$ group into an $SO(2N \!\!+\! 2)$ group and using the
$\!\frac{SO(2N \!+\! 2)}{U(N \!+\! 1)}\!$ coset variables
\cite{Fuk.77},
we have investigated a new aspect of the SUSY $\sigma$-model on
the K\"{a}hler manifold of the symmetric space
$\!\frac{SO(2N \!+\! 2)}{U(N \!+\! 1)}\!$.
A consistent theory of coupling of gauge- and matter-superfields to
the SUSY $\sigma$-model has been proposed on the K\"{a}hler coset space.
In the theory
a mathematical tool for constructing the Killing potential has been given.
Further we have applied the theory to the explicit construction of the SUSY $\sigma$-model on the coset space $\!\frac{SO(2N \!+\! 2)}{U(N \!+\! 1)}\!$.
We should emphasized again that
if one wants to develop some rigorous quantum-field theories
based on pure coset models,
inevitably one must face the very difficult problem of anomalies
in a holonomy group.
Such a problem particularly occurs in SUSY coset models
due to the existence of the chiral fermions
\cite{NNH.01}.
This is also the case for our
$\frac{SO(2N \!+\! 2)}{U( N \!+\! 1)}$
coset model
though the spinor rep of $SO(2N \!\!+\!\! 2)$ group is anomaly free.
But we were able to construct successfully the invariant Killing potential
in the present anomaly-free SUSY $\sigma$-model
which is equivalent to the so-called generalized density matrix
in the Hartree-Bogoliubov theory.
Its diagonal-block part is related to the present reduced scalar potential
with a Fayet-Ilipoulos term.

In order to see the behaviour of
the vacuum expectation value of $\sigma$-model fields,
after rescaling the Goldstone fields,
we have optimized the $f$-deformed reduced scalar potential
and found interesting $f$-deformed solution for an anomaly-free
$\frac{SO(2\cdot{\bf 5} + 2)}{SU({\bf 5} + 1) \times U(1)}$ SUSY $\sigma$-model.
The way of finding
these solutions, i.e.,
solutions for $Z^2$ (\ref{solutionforZ2f}) and
$< \!\!\chi_{\!f}\!\! >$ (\ref{solutionforchif}), is essentially different
from the way of finding the previous solutions in
(\ref{solutionforZ2chif})
which are quite the same as the ones obtained in I.
To observe clearly the difference in the solution forms,
we consider the special condition, $|{\bf k}|^2 \!=\! 1~(|k_0|^2 \!=\! 0)$,
and simplify a form of the solution for $Z^2$ as
\beqa
\BA{rl}
&\!\!\!\!
Z^2 
\!=\!
\left[ \!\!
\left[
{}^{^{^{^{^{^{^{.}}}}}}} \!\!
g^2 _{U(1)}
\!\cdot\!
\left\{ \!
4
\left( 1 \!-\! N \right)^2 \!
E
\!+\!
f^2 e^{f^2 {\cal K}_{f\sigma }} \!
\right\}
\!+\!
2 g^2 _{SU(N \!+\! 1)}
\!\cdot\!
2
\left[ \!
-
\left( 1 \!-\! N \right)^2 \!
E
\!-\!
{\displaystyle \frac{1}{4}}
f^2 
e^{f^2 {\cal K}_{f\sigma }} \!
{}^{^{^{^{^{^{^{.}}}}}}} \!\!
\right] \!\!
{}^{^{^{^{^{^{^{.}}}}}}} \!\!\!
\right] \!\!
\right] \\
\\[-8pt]
&\!\!\!\!
\times
\left[ \!\!
\left[
{}^{^{^{^{^{^{^{.}}}}}}} \!\!
g^2 _{U(1)}
\!\cdot\!
\left[
{}^{^{^{^{^{^{.}}}}}} \!\!
2 \left( 1 \!-\! N \right) \!
\left\{
f^2 \xi
\!+\!
\left(
\left( N \!+\! 1 \right) \! f^2 
\!+\! 12 \!-\! 14 N
\right) \!
E
\right\}
\right.
\right.
\right. \\
\\[-6pt]
&~~~~~~~~~~~~~
\left.
\!-
2 f^2
e^{^{^{\!f^2 {\cal K}_{f\sigma }}}} \!
N^2
\!+\!
{\displaystyle \frac{1}{2}}
f^2 e^{f^2 {\cal K}_{f\sigma }} \!
\left\{
8
\!+\!
\left( \!
f^2 \xi
{\displaystyle \frac{1}{E}}
\!+
\left( N \!+\! 1 \right) \! f^2 \!-\! 2 N \!
\right) \!
{\displaystyle \frac{1}{1 \!-\! N}}
\right\}
\right] \\
\\[-8pt]
&
+
2 g^2 _{SU(N \!+\! 1)}
\cdot
{\displaystyle \frac{1}{4}} \!
\left[ \!\!\!
{}^{^{^{^{^{^{^{.}}}}}}} \!
\left( N \!+\! 1 \right)
\left\{
\!-\!
2
\left(
3
\!-\!
f^2
\right)
\left( \! 1 \!-\! N \! \right)
\!+\!
8 \!
\left( 1 \!-\! N \right)^2
\right\} \!
E
\right. \\
\\[-6pt]
&~~~~~~~~~~~~~~~~~~~~~
+
16
e^{^{^{f^2 {\cal K}_{f\sigma }}}} \!
\left[
f^2 \!
\left( 1 \!-\! N^2 \right)
\!+\!
\left( 2 \!-\! f^2 \right)
{\displaystyle \frac{1}{1 \!-\! N}}
\right]
\!-
4 f^4
{\displaystyle \frac{N^2}
{\left( 1 \!-\! N \right)^2}}
e^{2 f^2 {\cal K}_{f\sigma }}
{\displaystyle \frac{1}{E}} \\
\\[-6pt]
&~~~~~~~~~~~~~~~~~~~~~
\!-
4
\left[
\left( 1 \!-\! N \right)
\left\{
\left( N \!+\! 1 \right) f^2
\!+\!
8 \!-\! 10 N
\right\} \!
E
\!+\!
e^{f^2 {\cal K}_{f\sigma }}
N^2
\right] \\
\\[-8pt]
&~~~~~~~~~~~~~~~~~~~~~
\left.
\left.
\left.
\!\!-
f^2
e^{f^2 {\cal K}_{f\sigma }}
\left[
{\displaystyle \frac{1}{1 \!-\! N}}
\left\{
\left( N \!+\! 1 \right) f^2
\!+\!
8 \!-\! 10 N
\right\}
\!+\!
e^{f^2 {\cal K}_{\!f\sigma }}
{\displaystyle
\frac{N^2}{1 \!-\! N}
}
{\displaystyle \frac{1}{E}}
\right]
{}^{^{^{^{^{^{^{.}}}}}}} \!\!\!
\right] \!
{}^{^{^{^{^{^{^{.}}}}}}} \!\!\!
\right] \!\!
\right]^{\!-1} .
\EA
\label{sol12VvariZf3}
\eeqa
Then the difference is evident.
The solution for $Z^2$ must be primarily positive
due to its square form and its positiveness should be analyzed.
Under the same condition,
a form of the solution for $< \! \chi_{\!f} \! >$ is also simplified as

\beqa
\!
\BA{rl}
&\!\!\!\!
< \! \chi_{\!f} \! >
=
\left( 1 \!-\! N \right) \!
\left[ \!\!
\left[
{}^{^{^{^{^{^{^{.}}}}}}} \!\!
g^2 _{U(1)}
\!\cdot\!
\left[
{}^{^{^{^{^{^{.}}}}}} \!\!
2 \left( \! 1 \!-\! N \! \right) \!
\left\{ \!
f^2 \xi
\!+\!
\left(
\left( \! N \!+\! 1 \! \right) \! f^2
\!+\!
10
\!-\! 12 N 
\right) \!
E
\right\}
\right.
\right.
\right. \\
\\[-10pt]
&~~~~~~~~~~~~~~~~~~~~~~~~~~
\left.
\!-
2 f^2
e^{^{^{f^2 {\cal K}_{f\sigma }}}} \!
N
\!+\!
{\displaystyle \frac{1}{2}}
f^2 e^{f^2 {\cal K}_{\!f\sigma }} \!
\left\{
6
\!+\!
\left(
f^2 \xi
{\displaystyle \frac{1}{E}}
\!+
\left( N \!+\! 1 \right) \! f^2 \!-\! 2 N
\right) \!
{\displaystyle \frac{1}{1 \!-\! N}}
\right\}
\right] \\
\\[-10pt]
&\!\!\!\!
+
2 g^2 _{SU\!(\!N \!+\! 1\!)}
\!\cdot\!
{\displaystyle \frac{1}{4}} \!
\left[ \!\!\!
{}^{^{^{^{^{^{^{.}}}}}}} \!\!
\left\{
\!-\!
2
\left(
3
\!-\!\!
f^2
\right) \!
\left( \! 1 \!\!-\!\! N^2 \! \right) \!
\!+\!
8 N \!
\left( \! 1 \!\!-\!\! N \! \! \right)^2
\right\} \!\!
E
\!-\!\!
4 \!
\left[ \!
\left(1 \!\!-\!\! N \right) \!
\left\{ \!
\left( N \!\!+\!\! 1 \right) \! f^2
\!\!+\!\!
8 \!-\! 10 N \!
\right\} \!\!
E
\!+\!
e^{f^2 {\cal K}_{f\sigma }} \!
N^2 \!
\right] 
\right. \\
\\[-8pt]
&~~~~~~~~~~~~~~~~~~
\!\!+
16
e^{^{^{f^2 {\cal K}_{f\sigma }}}} \!\!
\left[
f^2 \!
\left( 1 \!-\! N^2 \right) \!
\!+\!
\left( 2 \!-\! f^2 \right)
{\displaystyle \frac{1}{1 \!-\! N}}
\right]
\!-\!
2
f^2
e^{f^2 {\cal K}_{f\sigma }} \!
\!-\!
4 f^4 \!
{\displaystyle \frac{N^2}{\left( \! 1 \!-\! N \! \right)^2}}
e^{2 f^2 {\cal K}_{f\sigma }}
{\displaystyle \frac{1}{E}} \\
\\[-8pt]
&~~~~~~~~~~~~~~~~~~~~~~~~~~~~~~~~~~
\left.
\left.
\left.
\!\!-
f^2
e^{f^2 {\cal K}_{f\sigma }} \!
\left[
{\displaystyle \frac{1}{1 \!-\! N}} \!
\left\{
\left( N \!+\! 1 \right) \! f^2
\!+\!
8 \!-\! 10N
\right\}
\!+\!
e^{f^2 {\cal K}_{f\sigma }}
{\displaystyle
\frac{N^2}{1 \!-\! N }
}
{\displaystyle \frac{1}{E}}
\right]
\right]
\right] \!\!
\right] \\
\\[-10pt]
&~~~~~~~~~~~~~~~~
\!\times\!
\left[ \!\!
\left[
{}^{^{^{^{^{^{^{.}}}}}}} \!\!\!
g^2 _{U(1)}
\!\cdot\!
\left\{ \!
4
\left( \! 1 \!-\! N \! \right)^2 \!\!
E
\!+\!
f^2 e^{f^2 {\cal K}_{f\sigma }} \!
\right\}
\!+\!
2 g^2 _{SU\!(\!N \!+\! 1\!)}
\!\cdot\!
2 \!
\left[
-
\left( \! 1 \!\!-\!\! N \! \right)^2 \!\!
E
\!-\!
{\displaystyle \frac{1}{4}} \!
f^2 
e^{f^2 {\cal K}_{\!f\sigma }}
{}^{^{^{^{^{^{^{.}}}}}}} \!\!\!\!
\right] \!
{}^{^{^{^{^{^{^{.}}}}}}} \!\!\!
\right] \!\!
\right]^{\!-1} \!\! .
\EA
\label{solution2forchif3}
\eeqa
To obtain the solutions for $Z^2$
and
$< \!\!\chi_{\!f}\!\! >$,
we have imposed a condition
$
\left( \! 1 \!-\! N |{\bf k}|^2 \! \right) \!
\left( \! 1 \!-\! Z^2 \! \right)
\!=\!
|{\bf k}|^2 Z^2 \!\! < \!\!\chi_{\!f}\!\! > \!
(\! N \!=\! 5 \!)
$.
What does this condition mean?
It is an important and interesting problem to inquire upon
the physical meaning of the condition,
for example from the geometrical viewpoint.
Through a method different from the one in
\cite{NNH.01},
it is possible to determine
simultaneously a solution of the quadratic equation
for $E$ because it is given in terms of only
$Z^2$ and $< \!\!\chi_{\!f}\!\! >$.
We here omit an explicit expression for the equation
since it is very lengthy.
Particularly in the case $|{\bf k}|^2 \!=\! 1$,
the quadratic equation reduces to a simpler form.

In this paper,
along the same strategy developed by van Holten et al.
\cite{NNH.01},
we have embeded a coset coordinate
in an anomaly-free spinor rep of $SO(2N \!\!+\!\! 2)$ group
and have given a corresponding K\"{a}hler potential and
then a Killing potential for the anomaly-free
$\frac{SO(2N \!+\! 2)}{U(N \!+\! 1)}$ model
based on a positive chiral spinor rep.
The theory is invariant under a SUSY transformation
and the Killing potential is expressed in terms of the coset variables.
To construct a consistent gauged version of the SUSY coset model,
we must bring gauge fields into the model.
Then the theory becomes no longer invariant under the transformation.
To restore the SUSY, it is inevitable to introduce gauginos,
auxiliary fields and a Fayet-Ilipoulos term.
This makes the theory invariant under the SUSY transformation,
i.e., chiral invariant and
produces a new $f$-deformed reduced scalar potential.
Using such mathematical manipulation
we have thus constructed
the anomaly-free $\frac{SO(2N \!+\! 2)}{U(N \!+\! 1)}$
SUSY $\sigma$-model
and have investigated what are the new aspects
which had not been seen previously
in the SUSY $\sigma$-model on the K\"{a}hler coset space
$\frac{SO(2N)}{U(N)}$.


\newpage

\vskip0.8cm
\begin{center}
{\bf Acknowledgements}
\end{center}
S. N. would like to
express his sincere thanks to
Professor Manuel Fiolhais for kind and
warm hospitality extended to
him at the Centro de F\'\i sica Computacional,
Universidade de Coimbra, Portugal.
This work was supported by FCT (Portugal) under the project
CERN/FP/83505/2008.
The authors thank the Yukawa Institute for Theoretical Physics
at Kyoto University. Discussions during the YITP workshop
YITP-W-09-04 on ``Development of Quantum Field Theory and String Theory''
were useful to complete this work.

\newpage


\end{document}